\documentclass[manuscript,nonacm]{acmart}

\usepackage{hyperref}
\usepackage{hyperxmp}
\usepackage{amsmath}

\usepackage{placeins}

\AtBeginDocument{%
  \providecommand\BibTeX{{%
    \normalfont B\kern-0.5em{\scshape i\kern-0.25em b}\kern-0.8em\TeX}}}

\setcopyright{none}
\usepackage{ulem} 

\newcommand{\name}{our work}

\begin{document}

\title[ ]{Passive Mechanical Interfaces for Movement-Based Interaction}
\author{Jiawen Yao}
\affiliation{%
  \institution{University of Chicago}
  \city{Chicago}
  \state{Illinois}
  \country{USA}
}
\email{jiaweny@uchicago.edu}

\author{Yifan Zou}
\affiliation{%
  \institution{University of Chicago}
  \city{Chicago}
  \state{Illinois}
  \country{USA}
}
\email{angelzouyifan+pcs@gmail.com}

\author{Jiayu Fu}
\affiliation{%
  \institution{University of Chicago}
  \city{Chicago}
  \state{Illinois}
  \country{USA}
}
\email{fujiayu@uchicago.edu}

\author{Alan Pham}
\affiliation{%
  \institution{University of Chicago}
  \city{Chicago}
  \state{Illinois}
  \country{USA}
}
\email{apham799@uchicago.edu}

\author{Yue Jin}
\affiliation{%
  \institution{University of Chicago}
  \city{Chicago}
  \state{Illinois}
  \country{USA}
}

\author{Zhenglang Weng}
\affiliation{%
  \institution{University of Chicago}
  \city{Chicago}
  \state{Illinois}
  \country{USA}
}
\email{langlangw0202@gmail.com}

\author{Chi Wang}
\affiliation{%
  \institution{University of Chicago}
  \city{Chicago}
  \state{Illinois}
  \country{USA}
}
\email{chiwang@uchicago.edu}

\author{Liang He}
\affiliation{%
  \institution{University of Texas at Dallas}
  \city{Richardson}
  \state{Texas}
  \country{USA}
}
\email{liang.he@utdallas.edu}

\author{Ken Nakagaki}
\affiliation{%
  \institution{University of Chicago}
  \city{Chicago}
  \state{Illinois}
  \country{USA}
}
\email{knakagaki@uchicago.edu}


\renewcommand{\shortauthors}{Yao et al.}

\begin{abstract}
    Recent research in personal fabrication has enabled the creation of passive interactive materials; however, these works predominantly focus on contact-based input, such as pressing, bending, or twisting. We introduce \name{}, a mechanical construction strategy that extends this capability to inertial motion, enabling passive objects to sense and respond to movements like shaking, swinging, and tilting without external power.
By embedding mass within 3D-printed constraints, \name{} provides a modular system for responding to instantaneous, continuous, and state-based motions. To make these mechanisms customizable and accessible, we developed a parametric design tool that allows users to define desired motion behaviors and tune the underlying geometric parameters to generate corresponding inertial motion-enabled 3D models.
Finally, we demonstrate our approach through a range of electronics-free inertia-based interactive applications, from adding the affordance of everyday objects to supporting embodied activities.

\end{abstract}

\ccsdesc[500]{Human-centered computing~Systems and tools for interaction design}

\keywords{Interactive Material, Movement-based Interaction, Fabrication}

\begin{teaserfigure}
  \includegraphics[width=\textwidth]{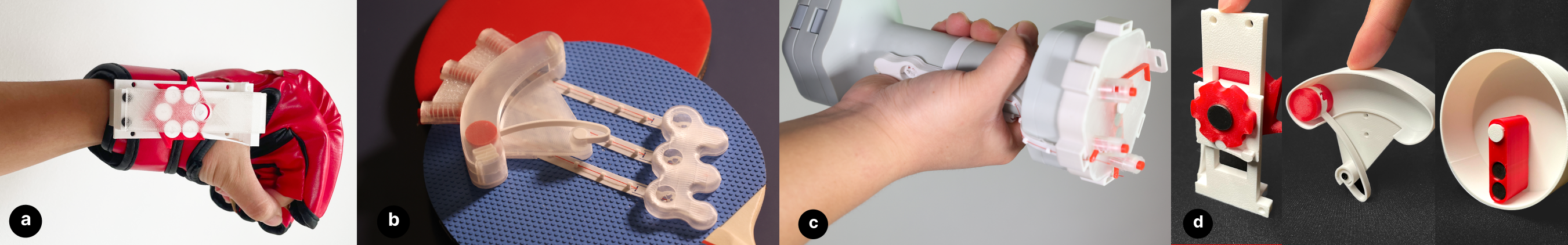}
  \caption{With (a) Punch repetition counting using the Stopper, (b) Table-tennis swing training using the Cantilever Beam, and (c) Dumbbell tilt-angle guidance using the Rotary Arm. (d) The three motion primitives provided by our work: Stopper, Cantilever Beam, and Rotary Arm. 
}
  \Description{Figure 1: "This image contains four major sections. From left to right, the first section demonstrates the application of our work for learning to swing in ping pong. The next section demonstrates a spray bottle opening specifically after a user has shaken it. The third section demonstrates a dumbbell with the added functionality of being aware of its tilted angle. The last section shows the developed GUI tool used for parametric design of the mechanisms."}
  \label{fig:teaser}
\end{teaserfigure}

\maketitle

\section{Introduction}

Movement is essential to numerous daily tasks. We shake, swing, tilt, rotate, and strike objects as part of routine activities, from interacting with tools to exercise and skill learning. Many of the artifacts involved in these activities are intentionally simple and passive: they require no power, introduce little technical overhead, and remain closely tied to their material and mechanical form. his motivates us to explore whether motion-responsive feedback can instead emerge from the passive mechanics of the object itself.

Recent work in HCI has shown that passive, battery-free materials and mechanisms can support sensing and feedback through physical form. However, most of these approaches remain centered on \textbf{contact-based motion}, where interaction is triggered through localized external forces (as defined in \cite{strongeImpactMechanics2018}) that deform the material, such as pressing, stretching, squeezing, or twisting \cite{zheng_shape-haptics_2022,mor_venous_2020,gong_metasense_2021,linFlexHapticsDesignMethod2022,kuriharaAugmentationKinestheticSensation2012}.
By contrast, many everyday activities rely on \textbf{inertia-driven motion}, where the object moves as a whole and its response depends on inertial dynamics rather than localized deformation \cite{strongeImpactMechanics2018}, such as when shaking a spray bottle, swinging a racket, punching, or lifting a dumbbell (Figure \ref{fig:contact_inertia}a). This raises a design question: \textit{How can everyday objects provide feedback to inertia-driven motion while preserving their passive nature?}

To explore this concept, we present \name{}, a passive mechanical system that allows mechanisms to respond to inertia- driven motion without relying on additional electronics. Enabling passive mechanisms to respond to inertia-driven motion is fundamentally challenging. While many \textit{contact-based} passive interactions exploit local force transfer and material deformation, \textit{inertia-driven} interactions require mechanisms to respond to motion of the object as a whole. To respond to inertia- driven motion, a mechanism must be sensitive to acceleration, impact, rotation, or momentum shifts distributed across the object. In \name{}, we leverage moving masses and mechanical constraints to translate whole-object motion into mechanically defined activation events, such as passing a mechanical gate, reaching positions along a constrained path, or reaching a target angular position. Tactile, visual, and mechanical output components can be added to provide perceptible feedback and close the interaction loop.
Our work also provides a parametric exploration tool that allows users to explore and compose the primitives, preview how geometric parameters shape their activation behaviors, and fabricate customized mechanisms as 3D-printable geometries.

In doing so, we open a design space in HCI for passive interactive mechanisms that can respond to inertia-driven motion through 3D-printable mechanical forms. Our contributions are as follows:

\begin{figure}[t]
    \centering
    \includegraphics[width=0.7\columnwidth]{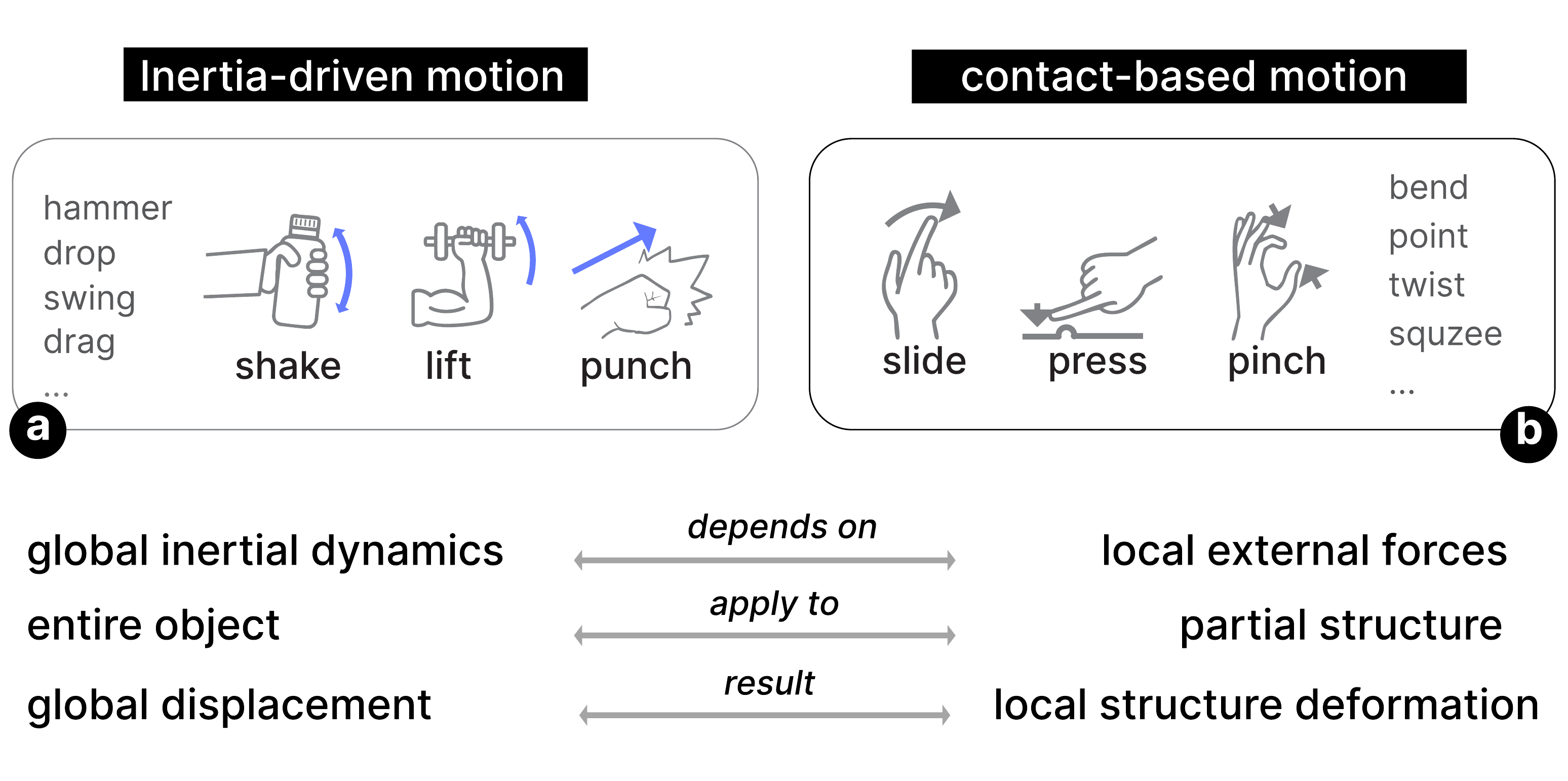}
    \caption{Comparison between (a) inertia-driven motion, in which global inertial dynamics act on the entire object and produce global displacement, and (b) contact-based motion, in which local external forces act on part of the structure and produce local deformation.}
    \label{fig:contact_inertia}
    \Description{Inertia-driven motion relies on whole-object inertial dynamics and produces global displacement, while contact-based motion relies on local external force and produces local deformation.}
\end{figure}

\begin{itemize}
    \item A mechanical system that can be composed to and reconfigured to respond to inertia-driven motion with perceivable feedback.
    \item A parametric tool that serves as an exploratory prototyping aid, enabling users to explore and customize primitive configurations and compositions by adjusting geometric parameters and previewing empirically derived motion references.
    \item A set of application examples demonstrating how the primitives can be combined with output modules to explore their potential across different objects and use scenarios.
\end{itemize}

\section{Related Work}
We build on prior work on motion-based 3D-printable devices and parametric design tools for customizing interactive behaviors. Extending this line of research, our work highlights customizable 3D-printable designs that use passive mechanisms to respond to inertia-driven motion. 

\subsection{Contact-based Input for 3D-Printable Interactivity}
To react to user input motion, a large group of work focuses on fabricating custom 3D-printed devices that interpret contact-based motion, which is quasi-static manipulation like touch, pressing, twisting, bending, and squeezing and requires continuous contact with the target device \cite{boemNonRigidHCIReview2019}, as interactive input \cite{zheng_shape-haptics_2022,he_squeezapulse_2017,changKirigamiHapticSwatches2020,ion_metamaterial_2016,alalawi_mechsense_2023,ogata_magneto-haptics_2018}. One common approach is to build custom haptic interfaces using 3D-printed compliant mechanisms \cite{zheng_shape-haptics_2022,changKirigamiHapticSwatches2020}. For example, \textit{FlexHaptics}~\cite{linFlexHapticsDesignMethod2022} provides a set of compliant beam-based modules that can be easily fabricated (e.g., using 3D printing or laser cutting) as building blocks for users to create custom haptic devices that respond to passive force input. Another approach explores interactive 3D-printed devices that take in user-input deformation behaviors like bending and squeezing for interactions \cite{yang_compact_2025, mor_venous_2020, ion_metamaterial_2016,lee_fluxlab_2026,he_squeezapulse_2017}. For example, \textit{LattiSense}~\cite{sakura_lattisense_2023} arranges conductive and non-conductive 3D-printed lattice-based structures to create custom deformation sensors, while \textit{DefSense}~\cite{bacher_defsense_16} embeds piezoresistive sensors into 3D-printed deformable input devices that continuously capture user-induced deformation by jointly optimizing sensor routing and signal-to-interaction mapping. 
These systems frequently pair contact with discrete or state feedback, such as depth haptic defined by the pressing displacement via compliant or bistable structures \cite{changKirigamiHapticSwatches2020,zheng_shape-haptics_2022}, or visual state changes (e.g., liquid mixing \cite{mor_venous_2020}). However, these designs, which respond to contact-based input, are difficult to support inertia-driven activities due to their dependency on structural deformation and localized forces.

\subsection{Inertia-driven Interactivity in 3D Printing}
Unlike contact-based input, inertia-driven motion arises when the user imparts energy and the device continues to move due to its mass and momentum (e.g., a push and flick), developing its subsequent behaviors with physical dynamics such as speed and acceleration \cite{strongeImpactMechanics2018}. Commercial Inertia Measurement Units (IMUs) have demonstrated programmable and robust abilities to recognize motion for interactions \cite{chung_onobject_2010,xu_hulamove_2021, zhou_imucoco_2025}. However, embedding off-the-shelf IMUs into 3D-printable devices requires dedicated mechanical designs and additional manual work to install the external hardware. HCI researchers have explored adding sensing capabilities to 3D-printed devices to support inertia-driven motion detection \cite{yang_recompfig_2022,gong_metasense_2021}. For example, \textit{MetaSense} \cite{gong_metasense_2021} integrates conductive materials into repeated cellular structure-based 3D-printed devices to perform capacitive sensing and detect acceleration. By 3D printing spring-based passive mechanical tags, \textit{MechanoBeat}\cite{oshim_mechanobeat_2020} encodes unique vibration frequencies for radar-based sensing, enabling battery-free identification and tracking of user interactions with objects. Finally, \textit{ReCompFig}~\cite{yang_recompfig_2022} combines compliant mechanisms with tensioned cables to create kinematic devices with dynamically reconfigurable degrees of freedom for detecting motion states. Unlike previous approaches, our work adopts a ``mass with constraint'' approach, inspired by commonly used motion-based mechanical devices such as tilt switches \cite{olson1941patent}, mechanical pedometers \cite{kato1982pedometer}, and spirit levels \cite{mehlman1964spiritlevel}, and applies magnets as the mass to extend the body of inertia-driven interactivity in 3D printing. Outside of HCI, a commercial product, ShockWatch \footnote{\url{https://www.shockwatch.com.au/}} has developed a stick-on attachment for tracking impact on shipping packages, inspired our research, while we intend to make such motion-activated mechanism customization via software and fabrication techniques.



\subsection{Fabrication Techniques and Design Tools for Motion-based Interactions}
To enable motion-based interactions with custom fabricated devices, researchers have developed techniques and tools that allow users to design and fabricate custom motion-driven interactive artifacts~\cite{ion_metamaterial_2016, gong_metasense_2021, he_ondule_2019, afsar_omnifiber_2021, he_kinergy_2022,savage_airlogic_2022}. For example, \textit{SqueezaPulse} \cite{he_squeezapulse_2017} provides a machine-learning-based pipeline, along with parameterizable corrugated 3D-printed tubes, to create distinct acoustic pulse profiles for interactivity. Similarly, \textit{FluxLab}~\cite{lee_fluxlab_2026} not only allows users to train a machine learning-based classifier, but also provides a parametric design tool for users to customize the deformable 3D model with integrated lattice design. Closest to our work, \textit{FlexHaptics}~\cite{linFlexHapticsDesignMethod2022}, \textit{Xstrings} \cite{li_xstrings_2025} and \textit{Kirigami Haptic Swatches}~\cite{changKirigamiHapticSwatches2020} offer parametric design tools that allow users to tune the target mechanism's geometric parameters to achieve desired user input detection.  In our work, we derive tunable physical properties for inertia-driven motion-related events from data-driven experiments and simulations, and we build parameter control into an interactive design tool that allows users to customize 3D-printable passive forms for desired inertia-driven motion detection.


Compared to prior work, our work views passive mechanisms as an interactive and customizable medium that can directly transform inertia-driven motion into interpretable states. Based on this perspective, we explore design opportunities for inertia-driven motion-based 3D-printable interaction through passive mechanisms that are discrete and tunable.

\label{sec:archi}
\section{ARCHITECTURE AND METHOD}
\begin{figure}[t]
    \centering
    \includegraphics[width=0.9\columnwidth]{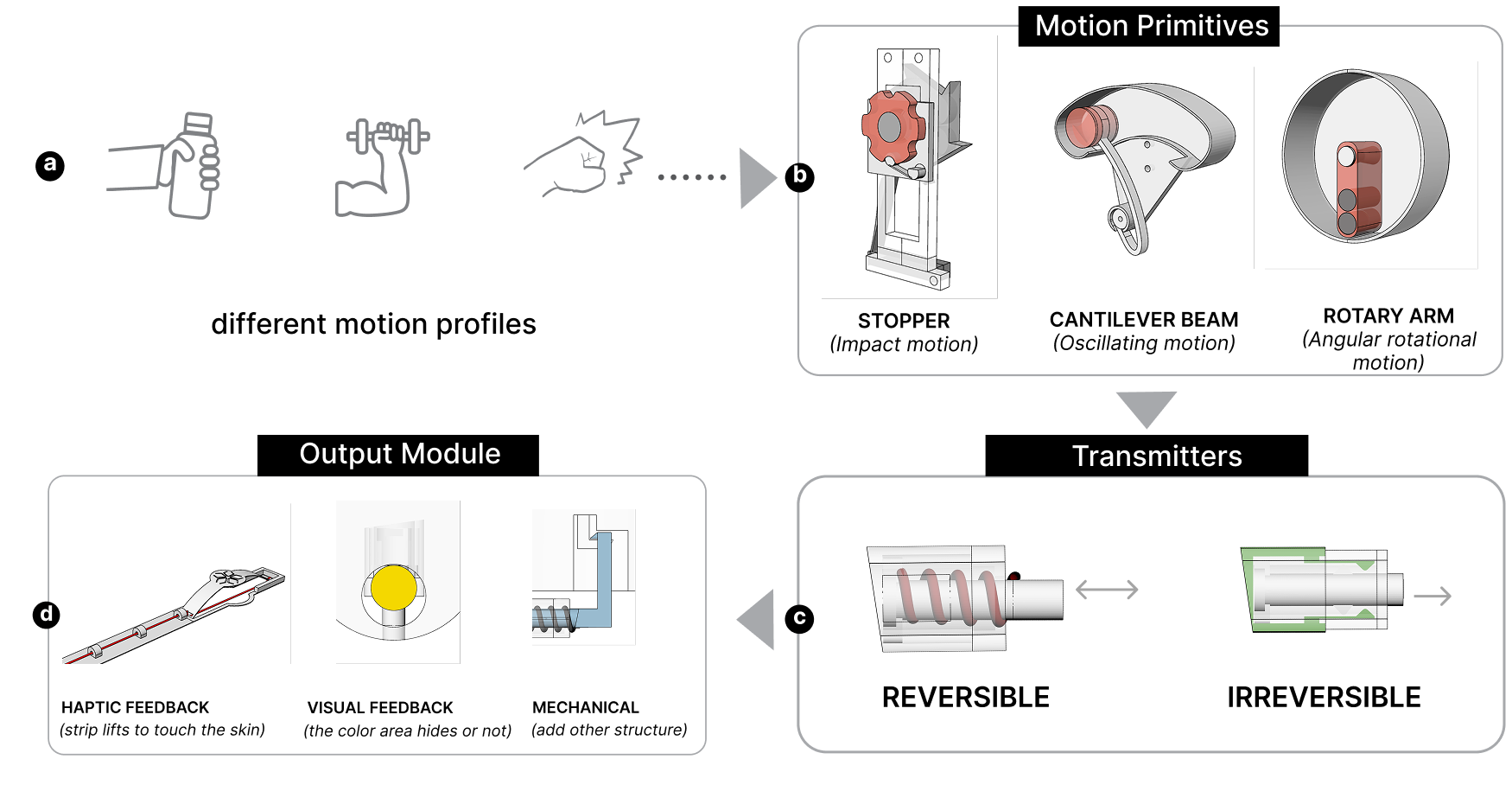}
    \caption{System overview. users begin with (a) a target motion pattern, select (b) a motion primitive, configure (c) a reversible or irreversible Transmitter, and choose (d) a tactile, visual, or mechanical output module.}
    \label{fig:architectrue}
    \Description{}
\end{figure}

Our work is a passive mechanical system that supports inertia-driven interaction from motion input to perceivable feedback by converting whole-object motion into discrete mechanical activation events. Figure~\ref{fig:architectrue} shows the overall system architecture. Rather than starting from a fixed taxonomy of motion~\cite{schmidtMotorControlLearning2019}, we developed three motion primitives, \textbf{Stopper}, \textbf{Cantilever Beam}, and \textbf{Rotary Arm}, as building blocks that respond to different motion patterns through distinct mechanical constraints (Figure~\ref{fig:architectrue}b). These primitives support abrupt motion, oscillatory motion, and rotational motion, respectively, and convert the relative motion of their internal masses into discrete activation events. We describe the design and behavior of each primitive in detail in Section 4.

\begin{figure*}[t]
    \centering
    \includegraphics[width=\textwidth]{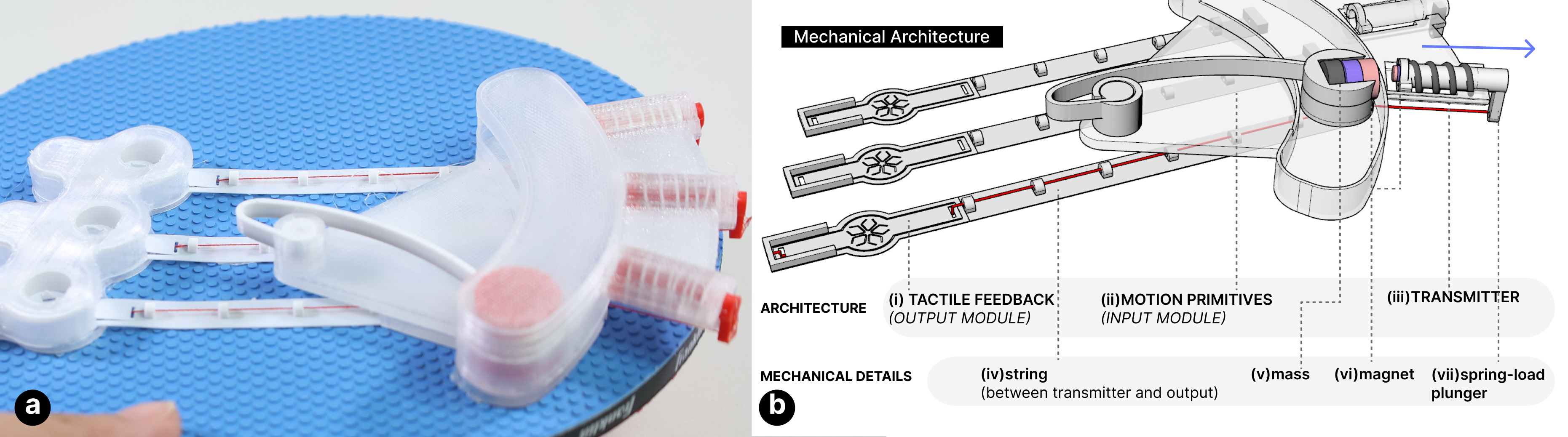}
    \caption{Physical prototype vs mechanical architecture. (a) A fabricated module with attached output modules. (b) The motion primitive, transmitter, and output module, together with their mechanical connections.}
    \label{fig:mech_structure}
    \Description{A two-part figure shows the assembled system and its mechanical architecture. The left panel shows a physical prototype with attached modules. The right panel shows the motion primitive, transmitter, output module, and key components, including the mass, embedded magnet, spring-loaded plunger, and connecting string.}
\end{figure*} To transform a primitive's mechanical activation event into perceivable feedback,users can combine transmitters and outputs with the three primitives (Figure \ref{fig:architectrue}c,d). Figure \ref{fig:mech_structure} shows one example of the connection between modules. Transmitters (Figure \ref{fig:mech_structure}b-iii) externalize the primitive’s internal state change as displacement, while outputs render that displacement as tactile, visual, or mechanical feedback (Figure \ref{fig:mech_structure}b-i). This separation allows users to choose the type of feedback they want to deliver and determine its placement on the object. Overall, \name{} consists of (1) motion primitives that convert inertia-driven motion into discrete mechanical activation events, (2) Transmitters that convert primitive's state changes into displacement, (3) output modules that transform displacement into perceivable feedback, and (4) a parametric tool that supports exploration of primitive configurations through empirically derived motion references.
In the following sections, we walk through the method and implementation of \name{}, including its mechanical architecture, empirical characterization, and parametric design tool.

\subsection{Method: Mass with Constraint}
\begin{figure}[htbp]
    \centering
    \includegraphics[width=1\columnwidth]{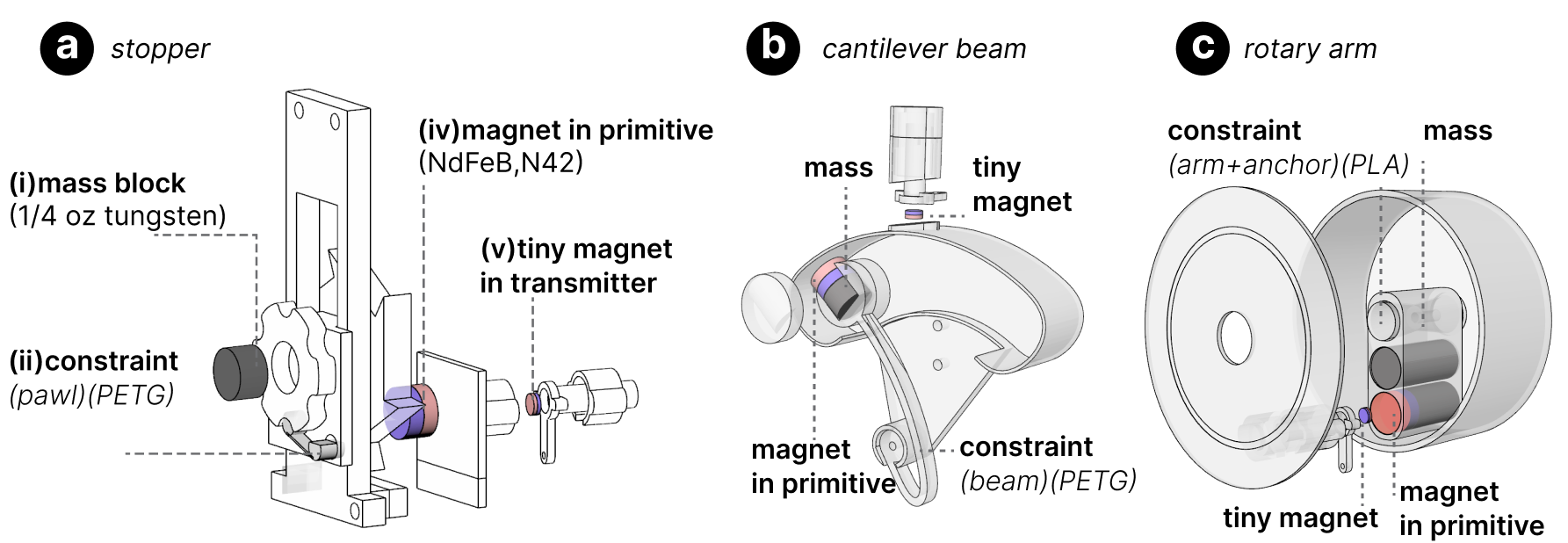}
    \caption{(a) view of the mass-with-constraint architecture. illustrated using the Stopper: (i) Tungsten mass block, (ii) Constraint types used across the three motion primitives, (iii) Magnet embedded in the motion primitive, and (iv) Magnet embedded in the transmitter. (b) overview of the mass-with-constraint for Cantilever beam. (c) overview of the mass-with-constraint for rotary arm}
    \label{fig:mass_with_constraint}
    \Description{An exploded view shows the mass-with-constraint structure and magnetic transmitter, including the mass block, constrained primitive, embedded magnet, and transmitter magnet.}
\end{figure}
In each motion primitive, we embed the dense tungsten masses(each is 1/4 oz) inside a 3D-printed structure that constrains its motion (Figure \ref{fig:mass_with_constraint}). Because the mass is heavier than the surrounding printed parts, it responds more strongly to inertia-driven motion. When the user moves the device, the mass exhibits relative motion within the constrained channel. The three primitives differ in the mechanical constraints they impose on the moving mass (Figure \ref{fig:mass_with_constraint}. These \textit{constraints} serve as mechanical filters that restrict the relative motion of the mass to specific directions and paths, shaping how different whole-object motion patterns produce discrete activation events. When the mass travels through the constrained channel and reaches the transmitter, the magnet (NdFeB N42) embedded in the primitive repels the tiny magnet inside the transmitter, causing the plunger to move and generate a displacement (Figure \ref{fig:mech_structure}b-vii) that can then drive the output module.

\subsection{Exploratory Parametric Tool}

\begin{figure}[b]
    \centering
    \includegraphics[width=1\columnwidth]{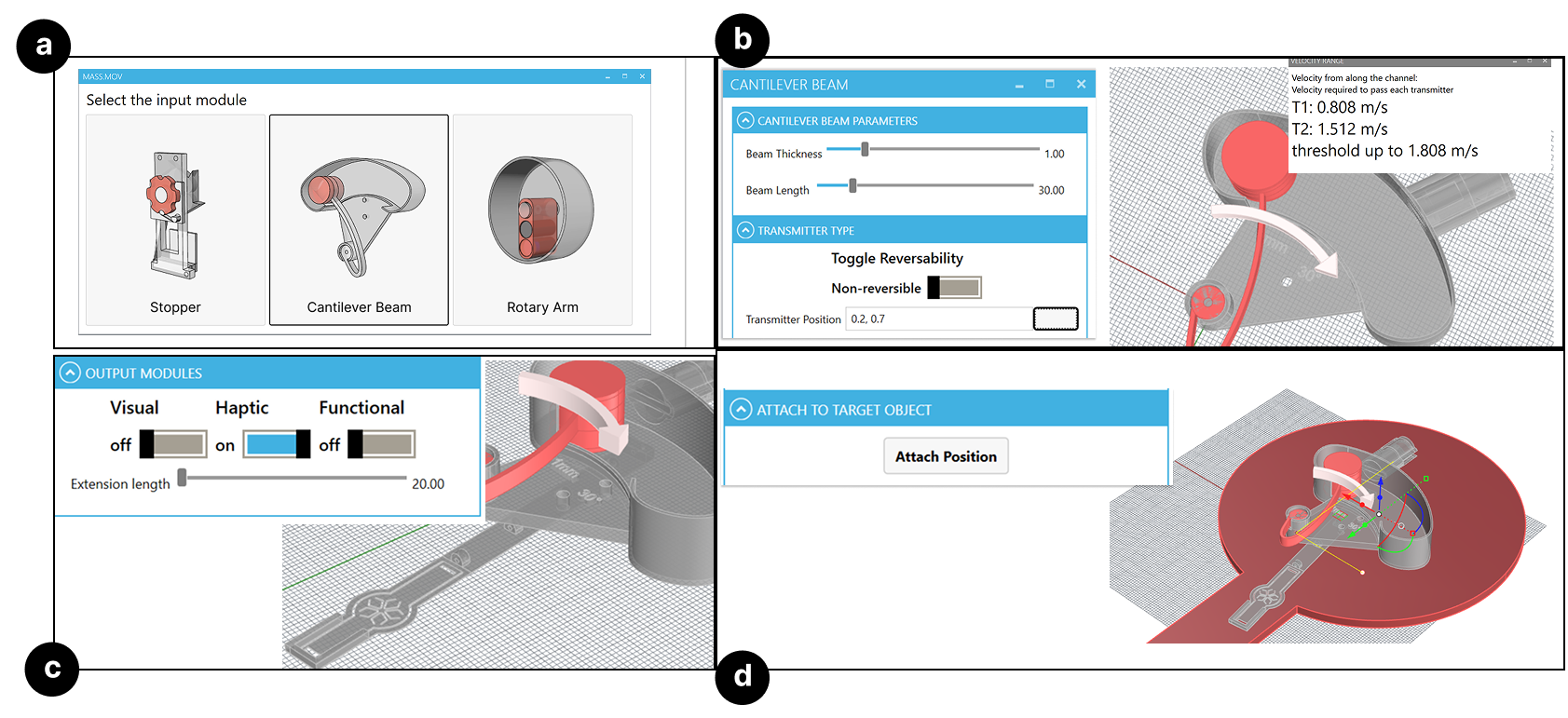}
    \caption{workflow in the parametric tool, from (a) selecting a motion primitive, (b) configuring its geometric parameters and motion reference (c) choosing output modules, and (d) mapping the module onto a target object }
    \label{fig:end_to_end_workflow}
    \Description{Three panels illustrate the end-to-end workflow in the parametric design tool. Panel a shows the selection of a motion primitive and the customization of its geometric parameters to explore different motion references. Panel b shows the configuration of the primitive and its mapping onto a target object. Panel c shows the placement of transmitters, the selection of output attachments, and the assembled module on a table-tennis paddle.}
\end{figure}

To help users develop their own applications, we developed a parametric design tool that supports the process of module configuration. The tool is intended as an exploratory interface that helps users examine different components, adjust their parameters, and compare configurations, rather than as a simulation tool for predicting exact mechanical behavior under real-world conditions. The user interface is implemented in Grasshopper within Rhino 7. It provides customizable parameters for all three motion primitives and supports connecting the primitives to Transmitters and output modules, as well as adapting the resulting configurations to target objects (Figure \ref{fig:end_to_end_workflow}).
Based on the system described in Section 3.1, we use a table-tennis paddle as an example to illustrate the workflow and how the modules are configured. In this example, the user wants to configure a Cantilever Beam for relatively faster swing motion and receive feedback when the mechanism reaches a selected activation position. The workflow is as follows (Figure \ref{fig:end_to_end_workflow}):

\begin{itemize}
    \item Select one of the three motion primitives.
    \item Customize the geometric parameters of the primitive while previewing the corresponding empirical motion reference.
    \item Configure the transmitters by specifying their positions relative to the input module.
    \item Choose an output attachment type and adjust its size.
    \item Import the 3D model of the target object (e.g., a table-tennis paddle in this case).
    \item Define the motion direction in the Rhino editing environment by drawing a curve, and map the customized module onto it.
    \item 3D print the module parts and assemble them to be attached to the real paddle. 
\end{itemize}Further details of the assembly and fabrication are provided in Appendix \ref{appendix:fabrication}.

\subsection{Empirical Design Reference}

\begin{figure}[htbp]
    \centering
    \includegraphics[width=1\columnwidth]{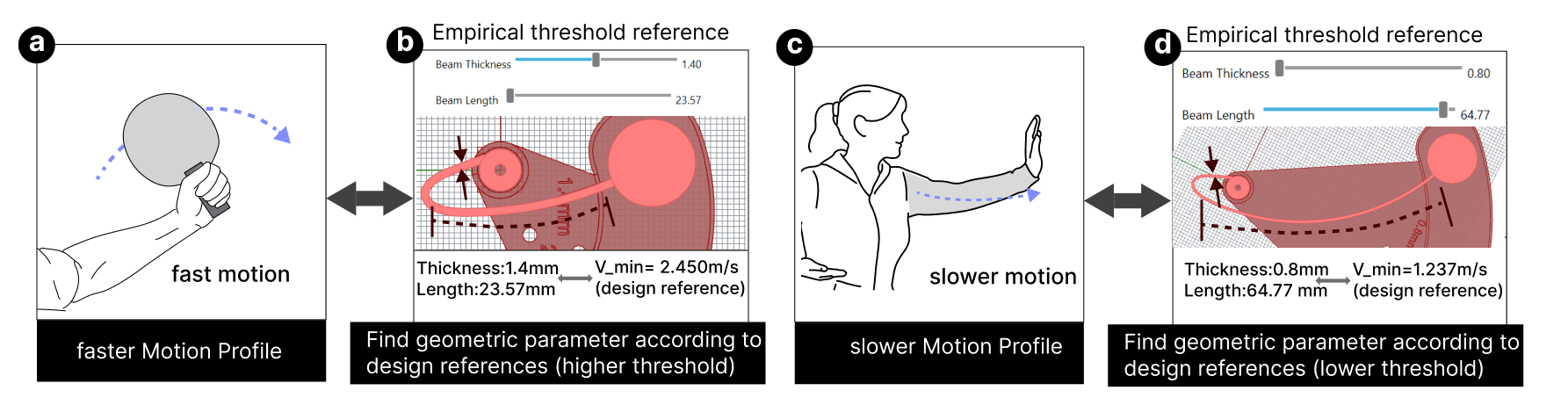}
    \caption{Configuring a Cantilever Beam motion threshold for table-tennis swing training. Panels (a) and (c) show target fast and slow motion profiles, while (b) and (d) show the corresponding beam geometries and empirically derived velocity reference.}
    \label{fig:architecture_threshold_to_geometry}
    \Description{}
\end{figure} To support design exploration, we empirically characterize how geometric parameters influence the activation behavior of each motion primitive. The mechanisms combine moving masses, compliant structures, magnetic interactions, and contact constraints, their responses are difficult to capture with a simple analytical model across different configurations. We therefore use controlled measurements and linear regression to establish quantitative motion references that relate geometry to activation behavior within the tested design space. This approach draws on the idea of \textit{surrogate modeling}, an engineering modeling approach that uses an empirical approximation to capture the relationship between selected design variables and system responses~\cite{gorissen_surrogate_2010}. Such models are useful when the underlying relationships among variables are too complex to describe with a simple analytical formulation. Similar empirical modeling approaches have also been adopted in HCI and digital fabrication research to relate fabrication or geometric parameters to resulting system behavior~\cite{linFlexHapticsDesignMethod2022,swaminathan_fiberwire_2019,changKirigamiHapticSwatches2020}.
In our work, these empirical references are intended to support comparison and configuration during design exploration rather than exact prediction under arbitrary real-world motion. Figure~\ref{fig:architecture_threshold_to_geometry} illustrates this process using the Cantilever Beam. By adjusting beam thickness and length, users can explore configurations associated with different derived velocity references. For example, a configuration with a higher velocity reference can be selected for a faster target motion profile, while a lower reference can be used when exploring slower motion.

\section{DESIGN SPACE AND CHARACTERIZATION}
Building on the architecture introduced in Section~\ref{sec:archi}, this section describes the design space of \name{} and characterizes its three motion primitives as building blocks, focusing on the types of motion patterns each primitive is suited for. We use three terms throughout this section. \textbf{Motion patterns} describe the temporal and directional forms of object motion of interest, such as brief motion bursts, oscillatory motion, and rotation. \textbf{Motion thresholds} describe the conditions under which these motions produce an activation event. \textbf{Geometric parameters} are the structural properties that users adjust to tune these activation conditions.

\medskip
\noindent\textbf{Design Rationale.}
Our work uses passive mechanical structures to threshold object motion and transform it into discrete, tunable activation \textbf{events}. 
The three primitives are distinguished by the motion patterns they are designed to respond to and by the mechanical behavior through which these motions produce an event. Rather than treating motion as a single physical quantity, we consider different temporal and directional patterns of object movement. Accordingly:\textbf{Stopper} responds to brief, abrupt motion bursts.
\textbf{Cantilever Beam} responds to oscillatory or repeated back-and-forth motion. \textbf{Rotary Arm} responds to rotational motion and angular position.

\medskip
\noindent\textbf{Geometric Parameters to Empirical Motion References.}
To help users understand how geometric parameters shape the motion response of each primitive, we empirically characterize their activation behavior under controlled motion conditions. Prior fabrication research has used a range of approaches to characterize the relationship between structural parameters and functional behavior, including passive haptic feedback~\cite{zheng_shape-haptics_2022}, sensing behavior~\cite{gong_metasense_2021}, and deformation and stiffness~\cite{li_xstrings_2025}. Such relationships can be modeled through simulation-based approaches such as finite element analysis (FEA), as well as empirical measurements and regression models~\cite{changKirigamiHapticSwatches2020}. We adopt an empirical characterization approach. These measurements provide motion references for comparing configurations and exploring the geometric design space, rather than predicting activation under arbitrary real-world motion. 
For each primitive, we characterize a motion-related quantity $Y$ as a function of its geometric parameters $\mathbf{P}$:
\[
Y_{\mathrm{ref}} = f(\mathbf{P}),
\]
where $Y_{\mathrm{ref}}$ represents a motion reference derived under our experimental setup, and $\mathbf{P}$ denotes the geometric parameter set. The user interface displays this reference as users adjust $\mathbf{P}$, allowing them to compare how different configurations shift the characterized response. We characterize the following relationships:
\begin{itemize}
    \item Stopper: reference deceleration $D_{\mathrm{min}}$, primarily varied by pawl thickness $T$.
    \item Cantilever Beam: reference deceleration $D_{\mathrm{min}}$ along the beam, varied by beam thickness $T$ and beam length $L$.
    \item Rotary Arm: angular resolution $\theta_{\mathrm{min}}$, determined by arm length $R$.
\end{itemize}

For the Stopper and Cantilever Beam, we fit empirical regression models that relate geometric parameters to the measured deceleration references under controlled motion conditions. For the Rotary Arm, the reference is derived from its geometric relationship between arm length and angular resolution. Detailed experimental settings are provided in the corresponding subsections. In the following subsections, we describe the design space of each primitive together with its empirical characterization.
\begin{figure}[t]
    \centering
    \includegraphics[width=1\textwidth]{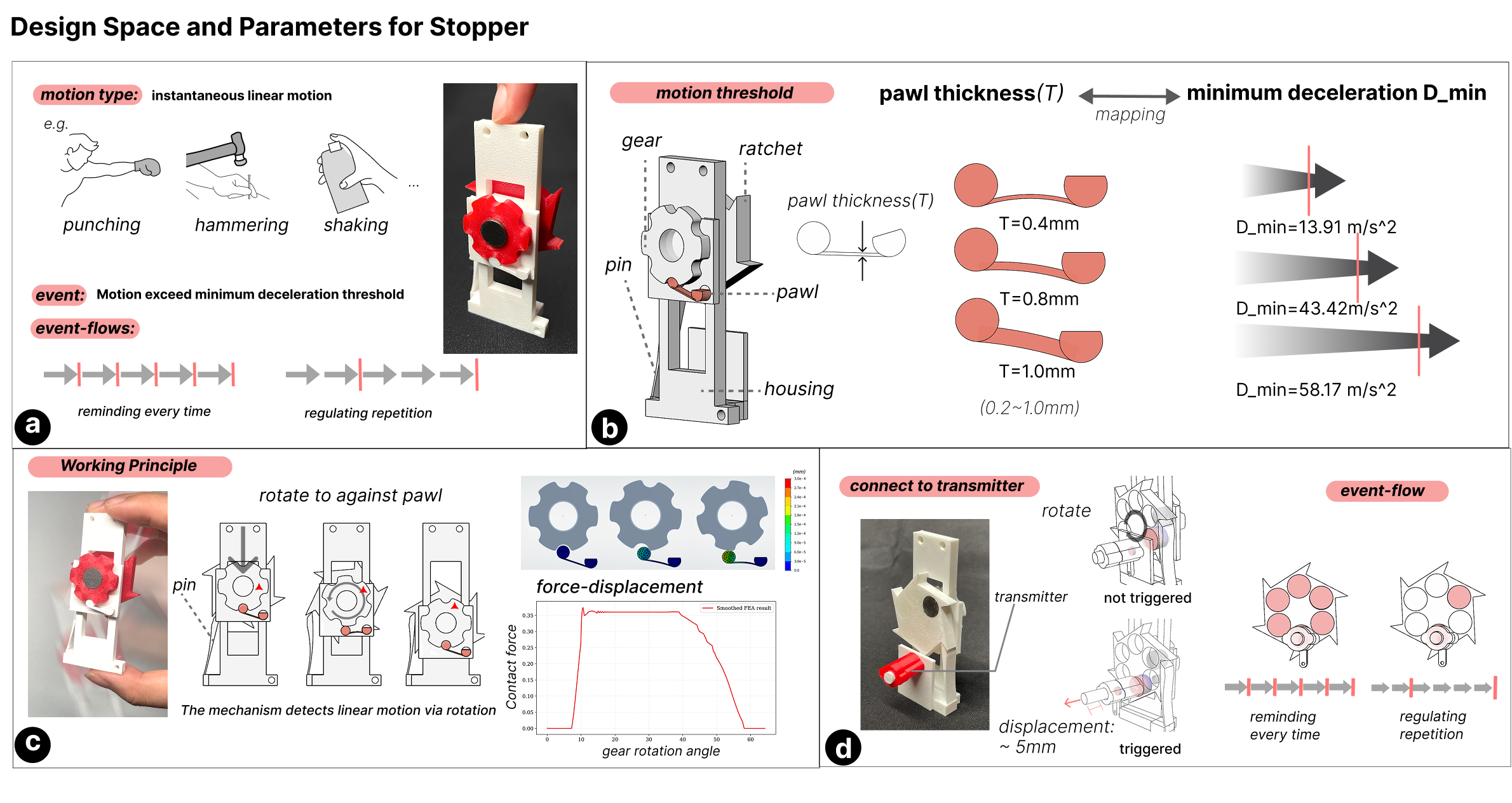}
    \caption{Stopper design space, including (a) target motion pattern and activation event, (b) mapping between geometric parameters and the empirical motion reference, (c) working principle, and (d) transmitter configurations for organizing triggered events.}
    \label{fig:stopper}
    \Description{A multi-part figure shows the design space and parameterization of the Stopper primitive. Panel a presents the target motion pattern and activation event for abrupt linear motion. Panel b shows how pawl thickness relates to the empirically characterized deceleration reference. Panel c illustrates the working principle of the mechanism. Panel d shows how the Stopper connects to transmitters to produce different event-flow configurations.}
\end{figure}

\subsection{Stopper for Abrupt Motion}

Stopper is designed for brief, abrupt motion bursts, such as punching, hitting, dropping, and hammering. During such motion, inertia drives the internal mass to move relative in the surrounding frame. An activation event occurs when this relative motion is sufficient for the mass to pass the mechanical gate formed by the pawl. Changing the pawl geometry therefore tunes how readily the mechanism is activated (Figure \ref{fig:stopper}c).

Beyond individual activation events, Stopper can be combined with the Transmitter and output modules to organize multiple activations into different feedback sequences, which we refer to as the \textit{event-flow}. Users can configure feedback to occur \textit{every time} the mechanism is activated, or only after accumulating a specified number of activations (e.g., every 2, 3, or 6 events), by arranging magnets at different positions on the Transmitter, as shown in Figure~\ref{fig:stopper}a and d. This supports interactions involving both immediate feedback for individual motion events and feedback based on repeated motion progress.

\medskip
\noindent\textbf{Characterization.} We use \textbf{deceleration} as the empirical motion reference for characterizing the Stopper's activation behavior. Specifically, we measure the deceleration associated with activation across different pawl geometries and use these measurements to characterize the minimum triggering deceleration $D_{\min}$ (Figure~\ref{fig:stopper}b). Inspired by a conventional ratchet mechanism~\cite{youtube_ratchet39}, the Stopper uses the pawl as a compliant mechanical gate. As shown in Figure~\ref{fig:stopper}c, the ratchet gear moves vertically within the housing. Under an abrupt motion input, inertia drives the gear downward, compressing a compliant pin that pushes the gear to rotate against the pawl. This converts the vertical relative motion into a rotational step of the ratchet. The activation event occurs only when the resulting motion is sufficient to overcome the mechanical resistance of the pawl. According to static force-displacement measurement (from FEA) the resisting force rises sharply as the gear pushes against the pawl, reaches a peak as it overcomes the pawl, and then decreases after passing it (Figure\ref{fig:stopper}c). We tune this resistance primarily through the pawl thickness $T$ (Figure~\ref{fig:stopper}b). A thicker pawl is stiffer and therefore generally requires a stronger motion input for activation. Rather than analytically deriving this relationship, we empirically characterize how pawl thickness shifts the deceleration associated with activation.

\medskip
\noindent\textbf{Apparatus.}
\begin{figure}[t]
    \centering
    \includegraphics[width=1\textwidth]{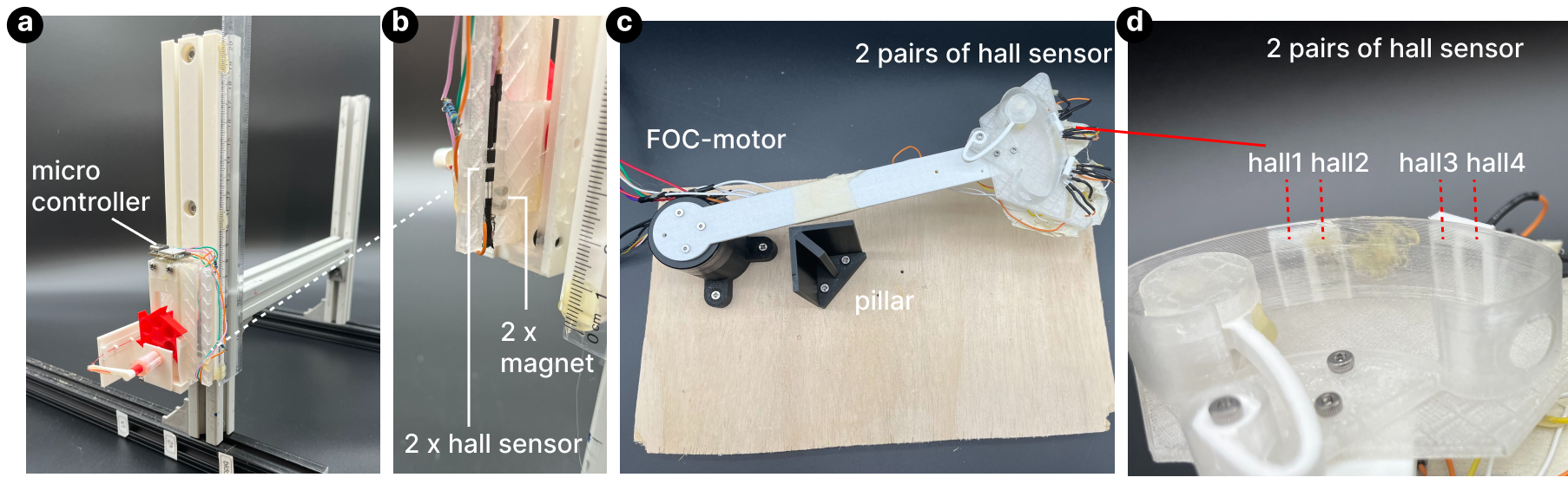}
    \caption{Experimental setups used to characterize the Stopper and Cantilever Beam for the linear regression models: (a) Stopper mounted on a vertical rail, (b) Hall sensor and magnet arrangement behind the Stopper, (c) Cantilever Beam mounted on the rotating test apparatus, and (d) two Hall sensor–magnet pairs positioned behind the Cantilever Beam.}
    \label{fig:set}
    \Description{}
\end{figure}
To characterize the triggering condition of each pawl-thickness configuration, we use a custom motor-controlled measurement apparatus (Figure~\ref{fig:set}a and b). Two magnets and two Hall sensors are used to measure the motion of the Stopper as the internal mass moves downward and attempts to overcome the resisting force of the pawl. To generate motion inputs of different magnitudes, we release the mechanism from different heights and identify the lowest release condition that produces activation for each configuration. The Hall sensors record the motion between two positions, from which we calculate the deceleration associated with the triggering event. We vary the pawl thickness $T$ across five configurations while holding the other geometric parameters constant:
\[
T \in \{0.2, 0.4, 0.6, 0.8, 1.0\}\,\mathrm{mm}.
\]
For each configuration, we use the minimum observed triggering deceleration as $D_{\min}$. 

\medskip
\noindent\textbf{Regression Model.} We fit a linear regression model to relate pawl thickness $T$ to the empirically measured minimum triggering deceleration $D_{\min}$:
\begin{equation}
D_{\min} = 73.76T - 15.59 + \Delta D_t + \Delta D_o
\end{equation}
The fitted model achieves an $R^2$ of $0.99$ within the tested parameter range. 
When the Stopper is connected to a Transmitter and an output module (Figure~\ref{fig:stopper}d), these additional components introduce further resistance to the mechanism. In our empirical model, we approximate their contributions as additive offsets, with $\Delta D_t = 0.12\,\mathrm{m/s^2}$ for the Transmitter and $\Delta D_o = 2.84\,\mathrm{m/s^2}$ for the output module. The resulting $D_{\min}$ is displayed in the user interface as a design reference while the user adjusts $T$.

\subsection{Cantilever Beam for Oscillatory Motion}
\begin{figure}[htbp]
    \centering
    \includegraphics[width=1\columnwidth]{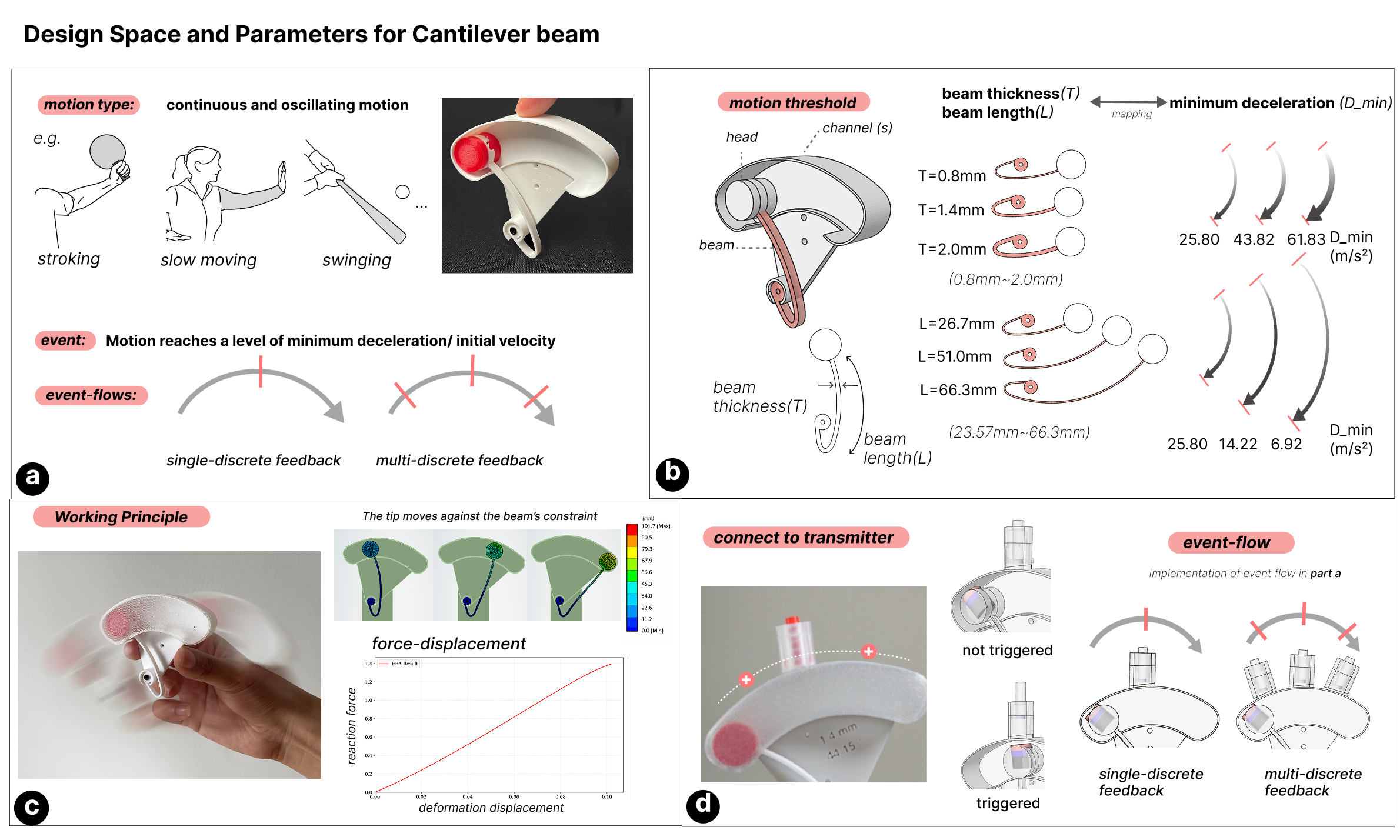}
    \caption{Cantilever beam's design space including (a) motion types, (b) parameter mapping, (c) working principle, (d) transmitter's configuration}
    \label{fig:cantilever_beam}
    \Description{A multi-part figure shows the design space and parameterization of the Stopper primitive. Panel a presents its motion type, event, and event for instantaneous linear motion. Panel b shows how pawl thickness maps to the minimum velocity threshold. Panel c shows the working principle of the mechanism. Panel d shows how the Stopper connects to a transmitter and supports different event logics.}
\end{figure}

Cantilever Beam is designed for continuous or oscillatory motion involving repeated changes in direction, such as swinging a paddle or performing repeated arm movements (Figure~\ref{fig:cantilever_beam}a). During user moving, inertia drives the cantilever head to travel in the channel. Unlike the Stopper, which produces a single gated transition, the Cantilever Beam provides multiple reachable positions along this path, allowing motion to be mapped to discrete feedback. As shown in Figure~\ref{fig:cantilever_beam}c, under static loading, the Cantilever Beam exhibits a relatively smooth and approximately linear increase in resisting force as the head is displaced along the channel, in contrast to the abrupt force change observed in the \textit{Stopper}.

With transmitters positioned along the channel, the primitive can support two event-flows (Figure~\ref{fig:cantilever_beam}d). A single transmitter can provide feedback when the cantilever reaches a selected position, while multiple transmitters can produce discrete feedback at successive positions along the channel. This supports interactions involving stepwise during repeated motion.

The motion of the cantilever head is shaped primarily by the beam geometry, particularly its thickness $T$ and beam arc length $L$ (Figure~\ref{fig:cantilever_beam}b). These parameters change the compliance and mechanical resistance of the beam as the head travels through the channel. A thicker or shorter beam is generally stiffer, whereas a thinner or longer beam is more compliant. The corresponding channel length $S$ also varies with the beam arc length $L$.

\medskip
\noindent\textbf{Characterization.} \textbf{Average deceleration} is used as an empirical motion reference for characterizing the Cantilever Beam, rather than as the physical quantity intrinsically sensed by the mechanism. This reference allows comparison of how different beam geometries shift the motion required for the cantilever head to travel through the channel. We characterize the average minimum deceleration associated with the cantilever head reaching the end of the channel, denoted as $D_{\mathrm{min}}$, across different combinations of beam thickness $T$ and beam arc length $L$.

\medskip
\noindent\textbf{Tested Configurations.} We tested three beam thicknesses,
$T \in \{0.8, 1.4, 2.0\}\,\mathrm{mm}$, and five beam arc lengths,
$L \in \{23.57, 34.20, 44.50, 55.62, 66.00\}\,\mathrm{mm}$.

\medskip
\noindent\textbf{Apparatus.} To measure $D_{\min}$ for each Cantilever Beam configuration, we built a custom motor-controlled measurement apparatus using field-oriented control (FOC), as shown in Figure~\ref{fig:set}c. The Cantilever Beam is mounted at the end of a rotating arm connected to the motor. Two pairs of Hall sensors are positioned near the beginning and end of the channel to measure the cantilever head's velocity as it enters and reaches the end of the channel (Figure~\ref{fig:set}d). During each trial, the motor drives the arm against a fixed pillar, producing a controlled abrupt change in motion that causes the cantilever head to travel through the channel. We gradually increase the commanded motor torque with $0.1\,\mathrm{N\cdot m}$ step and identify the lowest input at which the cantilever head reaches the end of the channel. The velocities measured by the two Hall-sensor pairs are then used to calculate the average deceleration associated with this triggering condition, which we use as $D_{\min}$ for that configuration.

\medskip
\noindent\textbf{Regression Model.} We fit a linear regression model relating beam geometry to the empirically measured minimum average deceleration $D_{\mathrm{min}}$:
\begin{equation}
\label{eq:cantilever_LR}
D_{\mathrm{min}}
=
-0.47L
+30.02T
+14.45
+\Delta D_t N_t
+\Delta D_o N_o .
\end{equation}
Here, $N_t$ and $N_o$ denote the numbers of connected Transmitters and output modules. In the fitted model, their contributions are approximated as additive offsets, with $\Delta D_t=3.73\,\mathrm{m/s^2}$ and $\Delta D_o=0.50\,\mathrm{m/s^2}$. The regression achieves an $R^2$ of $0.93$ within the tested parameter range.

\medskip
\noindent\textbf{Velocity Reference for the Interface.}
Although average deceleration provides a useful quantity for empirically characterizing the Cantilever Beam, it is less intuitive for users to interpret during design exploration. We therefore convert the empirically characterized minimum average deceleration $D_{\min}$ into a minimum initial velocity reference $v_{\min}$, which provides a intuitive motion reference for comparing configurations in the user interface. Here, $v_{\min}$ represents the estimated minimum initial velocity required for the cantilever head to reach a target position along the channel for a given beam configuration. Assuming constant average deceleration during the travel of the cantilever head, we use the kinematic relationship
\begin{equation}
    v_f^2 = v_i^2 - 2D_{\min}s,
\end{equation}
where $s$ is the travel distance of the cantilever head. At the minimum initial velocity required to reach the target position, the cantilever head reaches that position with $v_f=0$. Therefore,
\begin{equation}
    v_{\min}(s) = \sqrt{2D_{\min}s}.
\end{equation}
To make the effect of Transmitter placement similarly intuitive in the interface, we express the target position as
\begin{equation}
    s=nS,
\end{equation}
where $S$ is the total travel length of the channel and $n$ is the normalized Transmitter position, with $0<n\leq1$. Substituting this relationship gives
\begin{equation}
    v_{\min}(n) = \sqrt{2D_{\min}nS}.
\end{equation}
This formulation allows the interface to present the effects of both beam geometry and Transmitter placement through a single, intuitive velocity reference. As users adjust the beam thickness $T$, beam arc length $L$, or Transmitter position $n$, the interface updates the corresponding $v_{\min}$, helping users compare how different configurations may require relatively weaker or stronger motion inputs (Figure \ref{fig:end_to_end_workflow}b). We use this value as a design reference for exploration rather than as an exact prediction of activation velocity under arbitrary real-world motion.

\subsection{Rotary Arm for Angular Rotational Motion}

\begin{figure}[htbp]
    \centering
    \includegraphics[width=1\columnwidth]{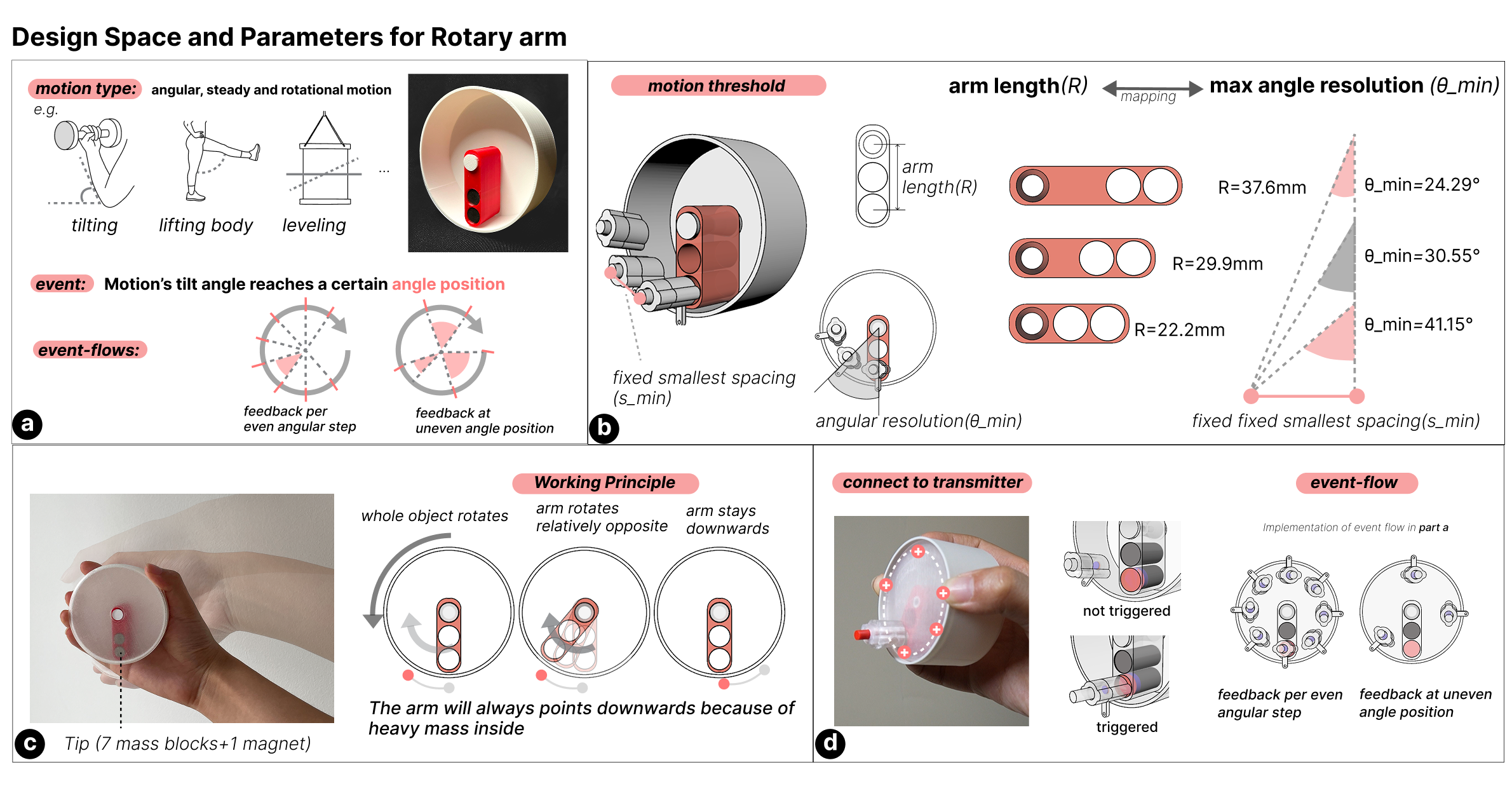}
    \caption{Rotary Arm design space, including (a) target motion patterns and event-flows, (b) geometric parameterization and angular resolution, (c) working principle, and (d) Transmitter configurations.}
    \label{fig:rotary_arm}
    \Description{A multi-part figure shows the design space and parameterization of the Rotary Arm primitive. Panel a presents rotational motion examples and supported event-flows. Panel b shows how arm length affects angular resolution. Panel c illustrates the working principle of the gravity-biased rotating arm. Panel d shows how Transmitters are positioned along the circular path to create discrete angular activation events.}
\end{figure}

Rotary Arm is designed for relatively steady rotational motion, where users are interested in angular position rather than abrupt motion changes (Figure~\ref{fig:rotary_arm}a). It converts continuous rotation into discrete angular position events. Users can configure two common event-flows (Figure~\ref{fig:rotary_arm}a and d): triggering repeatedly at evenly spaced angular intervals, or triggering at selected target angles. This supports interactions where users need to track progress through discrete angular positions or receive feedback when a target angle is reached.

The Rotary Arm uses \textbf{angular position} as its motion reference. To describe how finely the system can discretize rotation, we define the minimum angular step $\theta_{\min}$, which represents the angular resolution of the primitive. A smaller $\theta_{\min}$ corresponds to finer angular resolution.

The primary geometric parameter is the \textbf{arm length} $R$, defined as the distance from the rotation center to the magnet at the arm tip. Increasing $R$ increases the traveled arc length for the same angular displacement, allowing smaller angular steps to be distinguished (Figure~\ref{fig:rotary_arm}b).

As shown in Figure~\ref{fig:rotary_arm}c, the arm consists of eight overlapping mass blocks and a magnet near its tip. Gravity biases the arm downward, so the arm's orientation follows the rotation of the host object. Transmitters are placed along the circular path of the magnet. Each time the magnet passes a Transmitter, a discrete activation event is generated, dividing the continuous rotation into a set of angular states.

To determine a reliable minimum spacing between neighboring Transmitters, we performed a low-speed manual rotation test. The whole device of the Rotary Arm was rotated by hand under \(0.5~\mathrm{m/s^2}\) so that dynamic effects were minimized, allowing us to identify the smallest spacing at which adjacent Transmitters could still be activated separately rather than simultaneously. In this setup, the minimum reliable Transmitter spacing was
\[
S_{\min}=15.95\,\mathrm{mm}.
\]
We use this spacing as a geometric constraint for the current Rotary Arm design. Given this fixed spacing, the angular resolution is determined by the arm length $R$. Using the arc-length relationship
\[
S = R\theta,
\]
the minimum angular step is
\begin{equation}
    \theta_{\min} = \frac{S_{\min}}{R}.
\end{equation}
Thus, increasing $R$ decreases $\theta_{\min}$ and provides finer angular resolution.

\subsection{Transmitters}

\begin{figure}
    \centering
    \includegraphics[width=1\columnwidth]{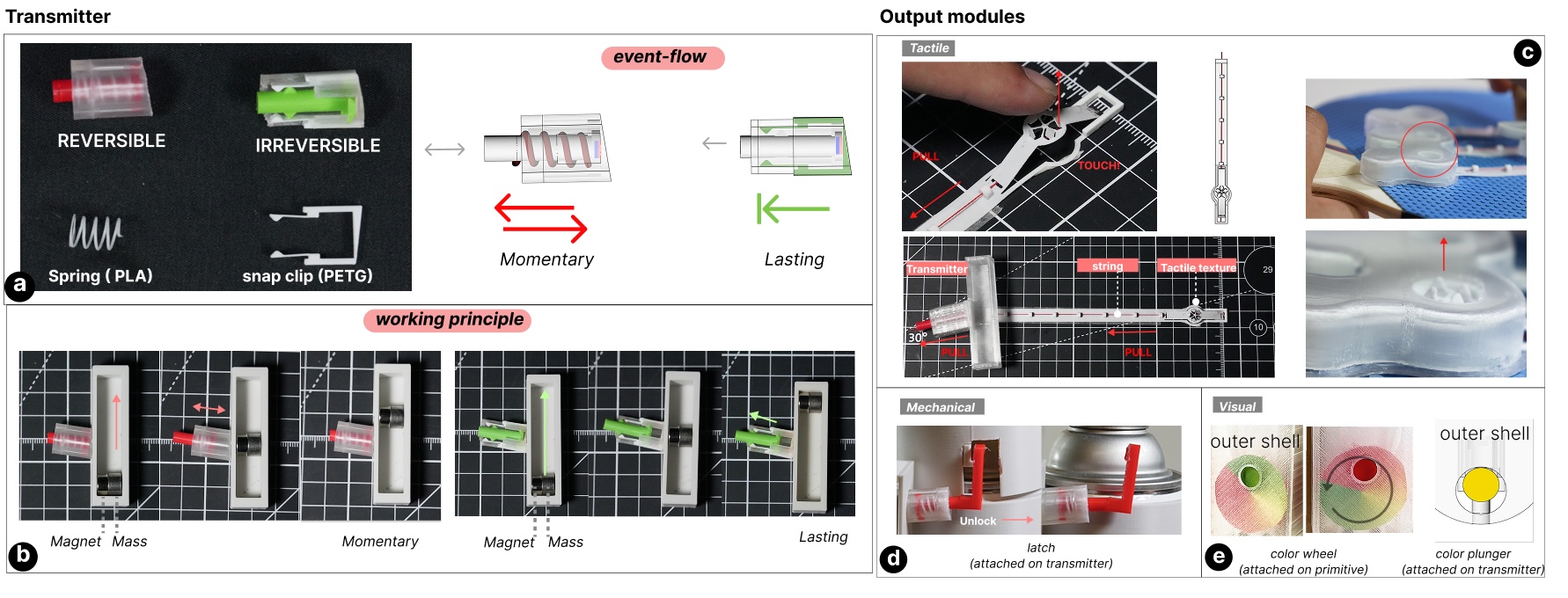}
    \caption{ (a) Reversible and irreversible transmitter designs and (b) their working principles. Output modules for different feedback
modalities. (c) tactile output touched to the skin, (d) mechanical output for latching and unlocking, and (e) visual output using
color-based state changes.}
    \label{fig:transmitter and out}
    \Description{Force measurement setups for the transmitter and tactile output.}
\end{figure}

To make the mechanical activation events generated by motion primitives perceivable to users, our work combines motion primitives with transmitters and output modules (Figure~\ref{fig:mech_structure}). The transmitter converts a primitive's internal state change into an external displacement that can drive downstream feedback modules (Figure~\ref{fig:transmitter and out}b), inspired by prior modular mechanical systems~\cite{nakagaki_trans-dock_2020,zhou_shape-kit_2025}. Importantly, transmitters also determine how an activation event is presented over time. We designed two types of transmitters to support either reversible or persistent responses (Figure~\ref{fig:transmitter and out}a). 

\begin{figure}
    \centering
    \includegraphics[width=1\columnwidth]{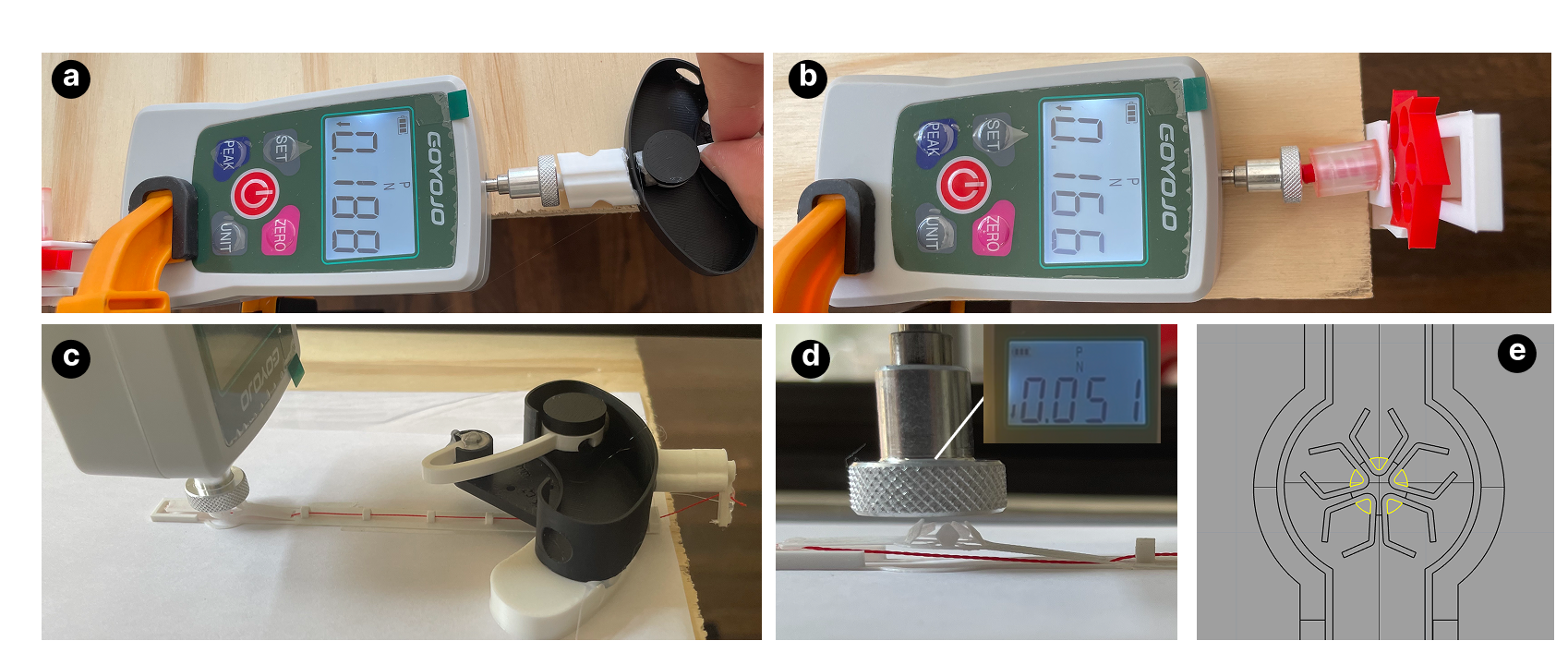}
    \caption{Force measurements for the reversible transmitter and tactile output.
    (a) Measurement setup for the cantilever beam with the attached transmitter, measured as 0.188 N.
    (b) Measurement setup for the stopper with the attached transmitter, measured as 0.166 N.
    (c) Measurement setup for the tactile output, transmitter, and cantilever beam.
    (d) The surface contact force of the tactile output was measured as 0.051 N.
    (e) Contact area of the tactile output calculated in Rhino: 5.38 mm$^2$.}
    \label{fig:force}
    \Description{Force measurement setups for the transmitter and tactile output.}
\end{figure}

\medskip
\noindent\textbf{Reversible transmitters.} The reversible one consist of a spring-loaded plunger and return to their initial state after actuation and can therefore be triggered repeatedly. When connected to the Cantilever Beam and Stopper, the reversible transmitter generated forces of approximately 0.188 N and 0.166 N as shown in Figure\ref{fig:force}a and b.

\noindent\textbf{Irreversible transmitters.} An irreversible transmitter contains a compliant latch that holds the plunger after actuation (Figure~\ref{fig:transmitter and out}a). When the magnet is activated to repel the plunger, the plunger passes through the latch and cannot return to its original position. This converts a transient activation event into a persistent mechanical state, allowing the result to remain visible or tangible until it is reset. After activation, the user can manually press the plunger to reset the transmitter.

These enable users to control whether feedback should respond to every successful event or only leave a mark. For example, although a Stopper mechanically triggers on every successful threshold crossing, an irreversible transmitter can be configured to react to only the first success, whereas a reversible transmitter can present feedback on every success.

\subsection{Output Modules}

Output modules take the displacement generated by the transmitter and convert it into perceivable feedback. They are designed around the form of feedback to be delivered, allowing different mechanisms to be attached downstream without changing the primitive's event flow. In this paper, we present three example output modules: tactile feedback, a mechanical latch, and visual feedback (Figure~\ref{fig:transmitter and out}). These examples demonstrate how the same transmitted displacement can be used to produce different feedback behaviors through different downstream mechanisms, building on prior work in mechanically mediated actuation and parameterized physical mechanisms~\cite{li_xstrings_2025,han_parametric_2023}.

\noindent\textbf{Tactile Feedback.} The tactile output produces a light touch on the skin through a textured surface actuated by the transmitter (Figure~\ref{fig:transmitter and out}c). The transmitter pulls a string connected to the textured surface, lifting it into contact with the skin. With a total contact area of 5.38~mm$^2$ (Figure~\ref{fig:force}), the estimated contact pressure was 9.48~kPa for the rigid condition and 8.55~kPa for the soft condition (a soft surface covered with a layer of hydrogel)

The design builds on \textit{Xstrings}~\cite{li_xstrings_2025}, which transfers mechanical motion to drive downstream behaviors in 3D-printed structures, and Han et al.'s parametric mechanism work~\cite{han_parametric_2023}, which demonstrates how physical mechanisms can be configured to produce different output responses.

Because the tactile output introduces additional mechanical load and resistance, it also affects the motion required to activate the system. This effect is therefore included as a parameter in the regression models for both the Stopper and Cantilever Beam.

\noindent\textbf{Mechanical Output.} The mechanical output uses a latch attached to the Transmitter to hold another component in a locked state (Figure~\ref{fig:transmitter and out}d). When the Transmitter is activated, its displacement moves the latch out of the locking position, releasing the constrained component. This release can then enable a mechanical state change in the attached object, such as opening, rotating, or releasing a spring-loaded part. In this way, the Transmitter displacement can be used directly as a mechanical input rather than being rendered as tactile or visual feedback. By attaching different latch-based mechanisms to the same displacement interface, \name{} can support a range of mechanically actuated outputs.

\noindent\textbf{Visual Feedback.} The visual output uses mechanical displacement to produce a visible color change. An outer shell conceals the colored element while leaving an opening at the position where the color is intended to appear. As the mechanism moves, the colored element enters or leaves this opening, revealing or hiding the color. This output can be implemented in two ways: by attaching a colored element to the moving part of the primitive, or by coloring the transmitter plunger so that its displacement changes the visible color state (Figure~\ref{fig:transmitter and out}e).

\begin{figure*}[htbp]
    \centering
    \includegraphics[width=\textwidth]{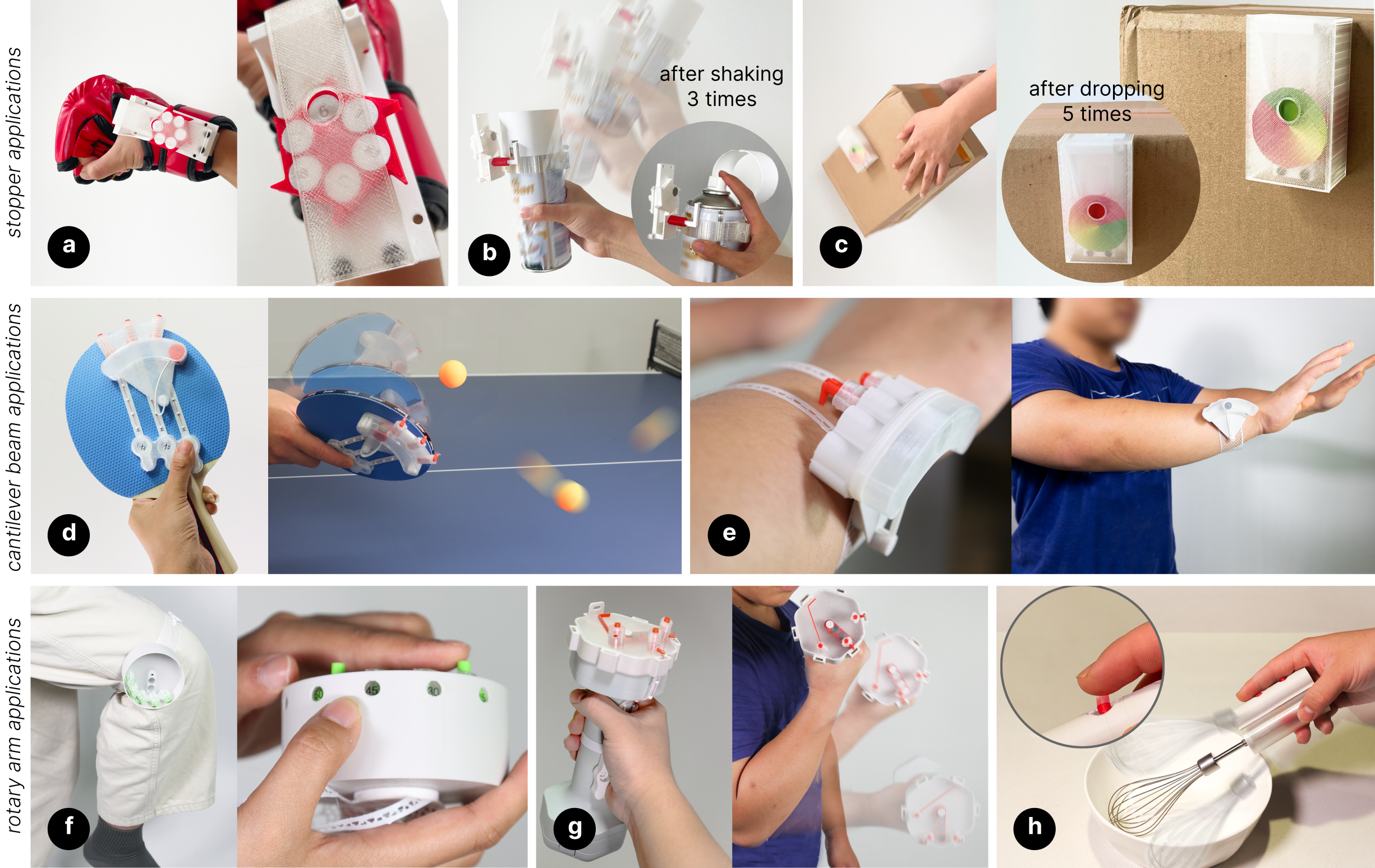}
    \caption{Application examples of \name{} across different motion primitives. Stopper: (a) punch counting, (b) shake-based access, (c) drop recording on shipped packages. Cantilever beam: (d) table-tennis swing feedback, (e) slow-movement training. Rotary arm:(f) arm-angle rehabilitation, (g) dumbbell lifting guidance, (h) whisking feedback.}
    \label{fig:applications}
    \Description{A multi-part figure shows application examples of \name{}. Panels a through c present Stopper applications: a boxing glove for counting punches, a spray bottle that unlocks after repeated shaking, and a shipping box that records dropping events. Panels d and e present Cantilever Beam applications: a table-tennis paddle that provides feedback on swing speed, and a wearable device for slow-movement training. Panels f through h present Rotary Arm applications: a wearable rehabilitation device for tracking arm angle, a dumbbell attachment for guiding lifting angle, and a whisk attachment that provides feedback on repeated whisking motions.}
\end{figure*}

\section{Applications}

We demonstrate a set of applications that use \name{} to support inertia-driven interactions across everyday objects and scenarios, grouped by the motion primitives they employ. These examples are intended to illustrate the breadth of the design space and the potential combinations of motion primitives and output modules, rather than to serve as in-depth evaluations of individual applications. As a system-level work, we use these applications to demonstrate the range of interactions enabled by \name{} and to motivate future application-specific exploration.

\subsection{Shake, Punch, and Drop with a Stopper}

Using the Stopper mechanism, \name{} can support interactions involving brief, abrupt motion events such as punching, shaking, and dropping. As shown in Figure~\ref{fig:applications}a, the mechanism can be integrated into a wearable object such as a boxing glove to register repeated punching events and display the accumulated count through a visual output module. Compared with an electronic sensing approach such as a smartwatch, this example illustrates how a passive, battery-free mechanism can provide physically configurable event-based feedback without additional sensing electronics.

When attached to everyday objects, the Stopper can also support repeated interactions in which multiple motion events are accumulated before an output is triggered. Figure~\ref{fig:applications}b illustrates this idea with a spray bottle, where the lid remains locked until the bottle has been shaken three times. After the required number of activations is accumulated, the mechanism releases a spring-loaded hinge and opens the lid. This interaction demonstrates how repeated inertia-driven events can be mechanically counted and mapped to a delayed output.

The Stopper can further provide a persistent record of motion events. As shown in Figure~\ref{fig:applications}c, when attached to a shipping package, the mechanism can retain a mechanical record of drop events that are sufficient to activate the Stopper. By adjusting the pawl thickness $T$, designers can tune how readily the Stopper is activated by a drop, while the event-flow configuration determines how many activations are retained or indicated. This application is inspired by commercial impact-history indicators used for monitoring potentially damaging handling during shipping.\footnote{\url{https://www.uline.com/BL_2639/ShockWatch-2-Indicators}}

\subsection{Swing and Slow Taichi Movements with a Cantilever Beam}

Using the Cantilever Beam, \name{} can provide feedback associated with different ranges of movement intensity or speed. Such a device can be attached to a table tennis paddle, providing tactile feedback at multiple positions along the beam path, which can be configured to correspond to progressively stronger swing motions (Figure \ref{fig:applications}d), and providing tactile feedback to the user's finger via output extensions. As the three haptic feedback modules are associated with three movement speeds, which are triggered at consecutive thresholds, the user can aim to activate a selected number of feedback points corresponding to the intended swing condition based on the target action (e.g., designers could associate different feedback levels with different table-tennis actions such as serving or blocking.).

Figure~\ref{fig:applications}e illustrates a potential wearable application for slow movement practice. By selecting a responsive Cantilever Beam configuration, the mechanism can be designed to activate under relatively smaller or slower body movements.

\subsection{Tilt, Lift and Whisk with a Rotary Arm}
The Rotary Arm can support applications involving angular-position feedback, including potential rehabilitation scenarios. For example, as shown in Figure \ref{fig:applications}f, \name{} can be attached to the body to to provide feedback at selected angular positions during repeated arm movements, such as raising the arm to a target angle each day. For example, feedback points could be positioned at selected target angles and adjusted over time to represent different movement goals. Similarly, a dumbbell can be augmented with a Rotary Arm to provide feedback on lifting angle, providing visual or haptic cues at selected lifting angles. Figure \ref{fig:applications}g shows a device that delivers haptic feedback at 60-degree increments, providing cues at $60^\circ$ increments throughout the lifting motion.

Finally, by embedding a Rotary Arm mechanism with a lead screw onto the handle of a whisk, the object could haptically and visually notify users when a preset number of whisking repetitions has been reached, responding to the user’s repeated whisking movements and providing feedback after a preset number of repetitions selected for a task such as whipping cream.

\section{USER STUDY}

 We conducted two user studies. The first examined whether the tactile feedback remained perceivable during body movement and whether participants found the feedback intuitive in relation to their motion. The second examined the usability of the exploration and composition workflow provided by our graphic user interface of parametric tool in Rhino 7. The study was approved by the IRB at our institution (IRB anonymous for review).

\subsection{Tactile Perception Study}
\begin{figure}[htbp]
\centering
\includegraphics[width=1\textwidth]{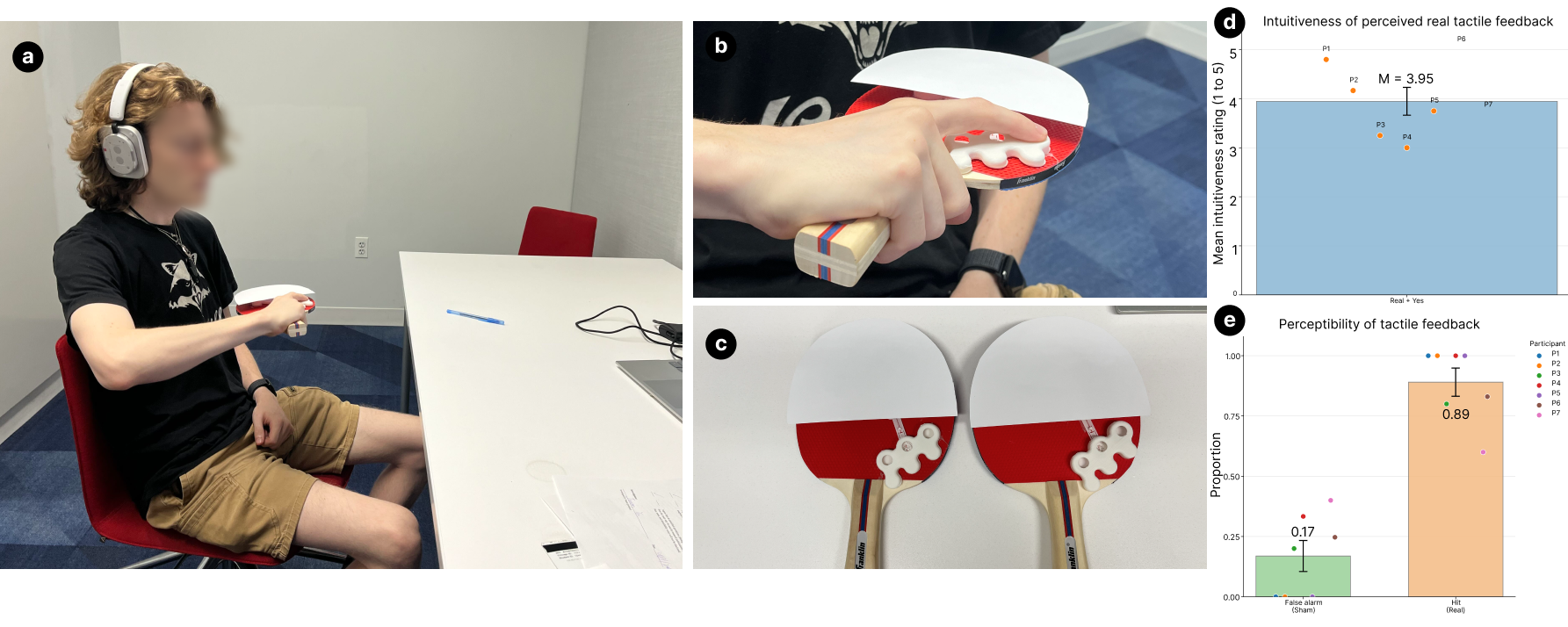}
\caption{Tactile perception study.
(a) Study setup. Participants wore headphones playing white noise while swinging the paddle.
(b) Finger position for receiving the tactile feedback.
(c) Sham and real conditions. The internal mechanism was covered to prevent participants from visually identifying the condition.
(d) Ratings of the intuitiveness of the tactile feedback in relation to participants' motion.
(e) Results of the perception check, showing participants' response accuracy.}
\label{fig:userstudy1}
\Description{}
\end{figure}
Our system has a modular composition system which supports tactile, visual, and mechanical output modules, together with a graphic user interface that allows users to explore, configure, and compose these components. In this session, we focus the perceptual evaluation on tactile feedback because, unlike visual or mechanical outputs that can be directly observed as state changes, tactile cues must remain perceivable while the user is actively moving the object and may therefore be more easily masked by motion-induced forces and vibration. 
To validate the tactile feedback from a user experience perspective, we conducted a perception study examining whether the feedback remained perceivable while participants were actively moving. We included both real and sham trials, presented in a randomized order to participants. After each trial, participants reported whether they perceived tactile feedback and rated how intuitive the feedback felt in relation to their motion. We invited 7 participants (mean age 22; 5 males and 2 females) to take part in the user study. 

\textbf{Procedure.}
The test device contained one tactile structure mounted on a ping-pong paddle. In each trial, participants swung the paddle three times. Before the formal study, we demonstrated the swinging motion and showed participants where to position their finger to receive the tactile feedback. During the study, participants wore headphones playing white noise and closed their eyes to reduce auditory and visual cues (Figure \ref{fig:userstudy1}a). For each trial, the researcher handed the participant one of two visually covered paddles corresponding to either the real or sham condition (Figure \ref{fig:userstudy1}c). The order of real and sham trials was varied across participants, and participants were not informed which condition they received. After each trial, participants answered two questions: (1) \textit{``Did you feel tactile feedback on your finger?''} and (2) \textit{``On a scale from 1--5, how intuitive did you find the tactile interaction in relation to your motion?''} At the end of the study, we also asked participants to describe the tactile feedback in their own words.

\textbf{Results.}
We calculated participants' response accuracy for distinguishing the real and sham conditions, as shown in Figure~\ref{fig:userstudy1}e. Overall, participants generally recognized the finger-based tactile feedback. We quantified perceptibility using the hit rate for real trials, defined as the proportion of real trials in which participants reported feeling the tactile feedback, and the false-alarm rate for sham trials, defined as the proportion of sham trials in which participants reported feeling feedback despite no finger-based tactile event being present. Across participants, the mean hit rate was 0.89, while the mean false-alarm rate was 0.17 (Figure~\ref{fig:userstudy1}e). Most false alarms occurred during sham trials, where participants sometimes interpreted the internal mass-induced weight shift of the paddle as tactile feedback at the finger. Among real trials in which the tactile feedback was perceived, participants also rated the interaction as relatively intuitive in relation to their swinging motion (M = 3.95/5), as shown in Figure~\ref{fig:userstudy1}d. Their verbal descriptions further illustrate how the tactile event was perceived. For example, P2 described it as \textit{``it feels like a ball is jumping on my finger,''} while P4 described it as \textit{``it feels like it strokes you.''}

\subsection{Parametric Tool Walkthrough Experience with Two Experts}

Our parametric tool also provides a graphic user interface (GUI) through which designers can explore the component design space and compose components according to their intended interaction. To examine the usability of this workflow, we invited two participants with professional-level expertise and training in CAD-related design fields to use our GUI in a design task (Table~\ref{tab:participants}). Participants were asked to complete the following task: \textit{design an interactive device for a golf-club-like object that could provide feedback related to its swing motion.}

We limited the task to the Cantilever Beam because it provides the largest number of configurable parameters among our three primitives and therefore represents a relatively complex case for examining the GUI workflow. During the study, we adopted a think-aloud method~\cite{eccles_think_2017}, encouraging participants to verbalize their thoughts, decisions, and feedback throughout the session. We provided a golf-club-like target object and an example in Rhino showing how a composed device could be attached to it. Participants were free to select the feedback output, adjust the component parameters, and determine the attachment location according to their own design rationale.

To help participants connect the parameters shown in the GUI with their physical effects, we also provided three pre-fabricated Cantilever Beam models with different configurations. Participants could handle and swing these models during the design process to experience their mechanical responses and feedback. The study therefore focused on how participants used the GUI together with motion references to explore and compose a design, rather than on evaluating the end-to-end fabrication process. 
\begin{table}[b]
    \centering
    \caption{Participant experience levels.}
    \label{tab:participants}
    \small
    \begin{tabular}{c p{0.75\linewidth}}
        \toprule
        \textbf{Participant} & \textbf{Experience Level} \\
        \midrule
        P1 & Proficient in Rhino and has used it extensively to develop design tools for creating customizable 3D-printable sensors and actuators, as well as other research prototypes and mechanical components. \\
        P2 & Three years of experience with mechanical structure design in Fusion 360, and one year of limited experience with Rhino. \\
        \bottomrule
    \end{tabular}
\end{table}

\begin{figure}[htbp]
\centering
\includegraphics[width=0.8\textwidth]{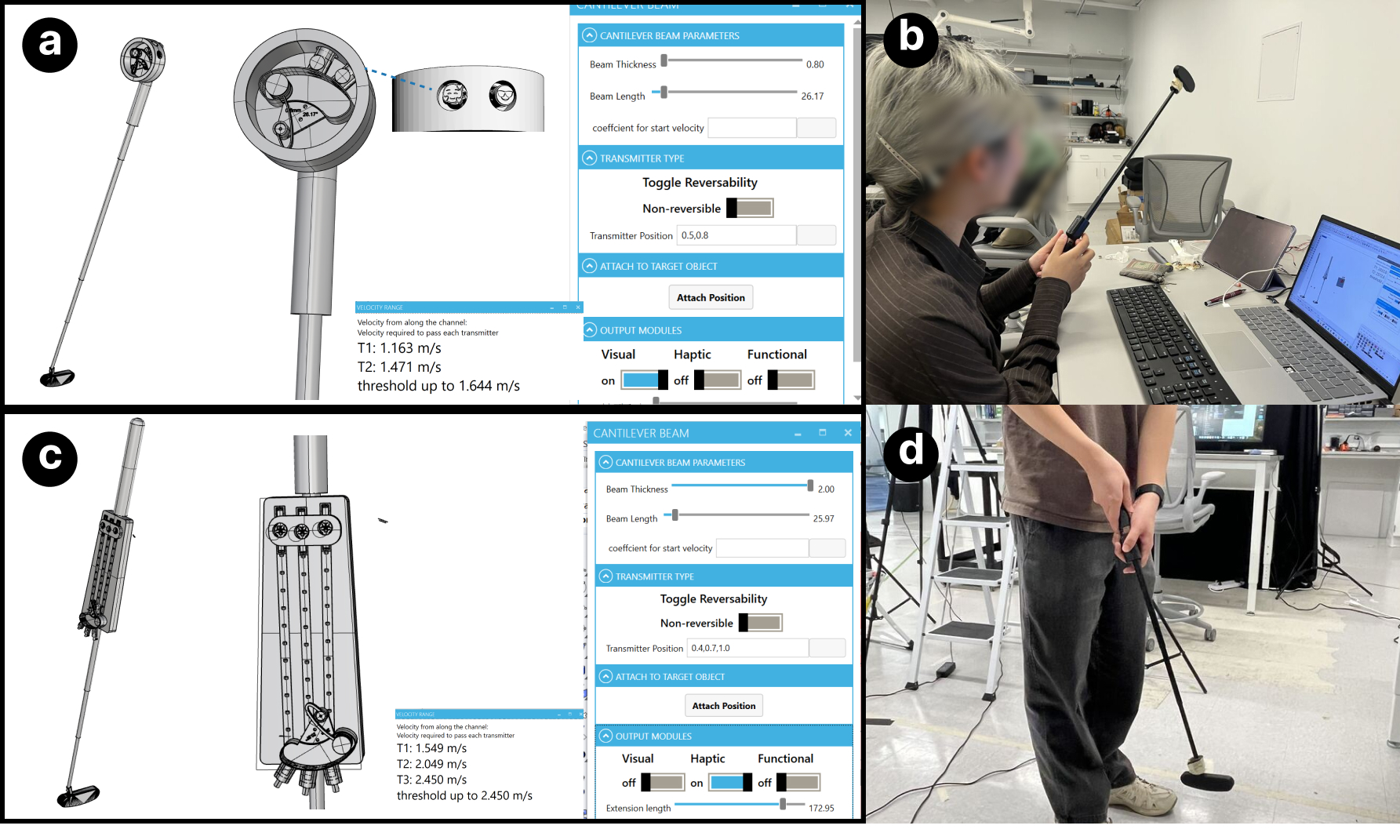}
\caption{Interface workflow walkthrough study.
(a) P1's design result, combining a visual output with the selected device configuration and parameters.
(b) P1 exploring where to anchor the device on the golf-club-like object.
(c) P2's design result using haptic outputs.
(d) P2 swinging the object to prototype and evaluate the configuration.}
\label{fig:userstudy 2}
\Description{}
\end{figure}

\textbf{Procedure.}
At the beginning of the study, we gave participants a walkthrough of the overall GUI We then asked them to physically try three example cantilever configurations: $L=23.57$~mm, $t=2$~mm; $L=44.5$~mm, $t=1.4$~mm; and $L=66$~mm, $t=0.8$~mm. These examples represented different points in the cantilever design space. Participants swung each configuration and explored the approximate motion threshold required to trigger its output. This familiarization step allowed them to connect the physical behavior of the component with the reference thresholds presented in the GUI. Participants then completed the design task using the GUI, which consisted of two parts: (1) composing the device by selecting an output type and configuring its parameters, and (2) designing the connecting geometry needed to attach the resulting device to the golf-club-like object.

\textbf{Results.}
The two participants developed different design strategies for providing feedback during a golf swing, as shown in Figure~\ref{fig:userstudy 2}a. P1 had prior-experience with golf and selected visual feedback as the output. She placed the device near the handle and explained that haptic feedback might be difficult to notice during a highly focused and fast movement. Instead, she proposed using two Emoji symbols to distinguish between a successful swing and one that was close to the desired performance (Figure ~\ref{fig:userstudy 2}a). To integrate the device, she created a circular attachment around the handle. For the mechanical configuration, P1 wanted the device to remain relatively small while responding to a moderate swing speed. She therefore selected $L=26.17$~mm and $t=0.8$~mm, with transmitters positioned at approximately 0.5 and 0.8 along the cantilever. The GUI showed corresponding reference thresholds of approximately 1050.2~mm/s and 1328.4~mm/s.

P2 instead selected three haptic outputs and positioned the device below the handle so that the outputs could contact the fingers while holding the golf club. P2 explained that, because they had no prior experience with golf, they expected to focus primarily on the club head and the overall swing motion. They therefore considered finger-based haptic feedback useful because it could provide information without requiring additional visual attention. After trying the sample configurations, P2 observed that the golf-club-like motion involved relatively high swing speeds. They therefore selected a Cantilever Beam configuration associated with a higher and broader range of motion references, using $L=25.97$~mm and $T=2$~mm, with Transmitters positioned at approximately 0.4, 0.7, and 1.0 along the cantilever path, as shown in Figure\ref{fig:userstudy 2}c.

Participants were also asked to complete a post-study questionnaire rating their experience with the GUI on a 7-point Likert scale. Both participants rated all four aspects of the workflow positively (5--7 out of 7) with mean scores of 6.0 for translating movement into parameters, 6.0 for understanding parameter–behavior relationships, 6.0 for mechanism integration, and 6.5 for confidence in design decisions. 

\textbf{Take-away.} 
Overall, the study showed that participants could use the GUI to explore parameters and develop their own designs, however, the velocity reference was initially somewhat confusing, as participants were unsure what level of movement speed the displayed values represented in practice. After trying the provided physical samples, they developed a clearer understanding of the relative trend of the references. For example, they recognized that thicker mechanisms generally corresponded to higher movement velocities, and used the thicker sample as a reference for the higher end of the velocity range and the thinner sample for the lower end. P1 noted that seeing a sample mechanism integrated into the golf club could help designers better understand how to select parameters for a specific object. This suggests that providing both parameter-based guidance and template examples can make the design process easier to understand. In future work, we would like to extend this into a complete end-to-end workflow by providing sample models, object-specific examples, or fabrication-ready templates to better connect configuration with physical prototyping.

\section{Discussion, Limitations, and Future Work}

While \name{} demonstrates that passive objects can provide perceivable feedback on inertia-driven motion through mechanically structured modules, it also reveals several limitations that point to broader directions for future work.

\label{sec:discussion}
\subsection{Integrating and Composing Modules}

At present, the modules are still better suited as attachable components than as structures fully integrated into the artifacts they augment. Their current size, dependence on flat or smooth mounting surfaces, and reliance on compliant printed parts limit seamless deployment in everyday objects. Future work could address this through higher-resolution fabrication, more robust material choices, object-specific adapters, and possibly co-fabricating the mechanism directly with a target 3D object. Smaller and denser fabrication approaches, such as metal 3D printing technologies like Desktop Metal\footnote{Desktop Metal. \url{https://www.desktopmetal.com/}}, may further help reduce the overall footprint of the modules while preserving their mechanical function. Also, to best integrate the module into everyday objects and tools, miniaturizing the mechanisms need further work, and relying on existing metal spring products or other miniaturized mechanisms could be another way. 

Connected modules also introduce additional complexity. While our characterization already incorporates the effects of transmitters and outputs, it currently assumes that their contribution to threshold behavior remains consistent across configurations. In practice, however, threshold values may shift with different linkage directions, attachment styles, and assembly relationships. More configuration-sensitive models could therefore better support composed assemblies in both analysis and design tools.

\subsection{Challenges in dynamic characterization}
While we provide motion references to help users select appropriate parameters, these values are empirically derived under controlled experimental conditions and should not be treated as universal thresholds. As with empirical surrogate models more broadly, predictions are most reliable within the characterized design and operating range, while extrapolation to unseen conditions can introduce additional uncertainty~\cite{gorissen_surrogate_2010,anastasio_investigation_2023}. 

For example, the Cantilever Beam was characterized with the mechanism anchored horizontally, while the Stopper was characterized under a vertical deceleration condition. In real-world use, however, the orientation and mounting position of the mechanism may vary. The internal mass may move with gravity, against gravity, or along an intermediate direction, which can change the dynamic response of the mechanism and the motion required to trigger it. Our current characterization also focuses on relatively isolated motion events, asking how much velocity or deceleration is required to overcome the mechanical resistance and produce an activation. More complex motions may therefore behave differently.

\subsection{Adapting the motion reference to actual movement}

We describe our measurements as empirical motion references, they capture useful trends across geometric configurations, rather than providing a single threshold that applies to every object, orientation, or movement condition. One direction for future work is to make these references adaptive to the user's actual motion. We used to include an re-calibration function in GUI that collects motion data from the user and adjusts the predicted threshold using a calibration coefficient. However, we found that real user motion is often less controlled than our bench experiments. For example, users may shake an object repeatedly in both directions, producing motion traces that contain multiple accelerations, reversals, and activation events rather than a single clean input. Future systems could therefore use learning methods, such as \textit{Bayesian algorithm}~\cite{snoek_practical_2012}, to continuously update the motion model as new interaction data are collected. This could allow the system to gradually adapt its motion references to different mounting orientations, objects, and individual movement patterns.

\subsection{Expanding non-contact interactive mechanism system }

These limitations point beyond the current module set. Future work could explore how multiple modules might be combined on a single object to support layered interpretations of motion, simple mechanical logic, or the superposition of multiple dimensions of motion. Such systems might also extend across multiple objects and support multimodal feedback. For example, in practicing latte art, a Rotary Arm module could be attached to the milk-pitcher to indicate its tilt, while a Cantilever Beam module on the pitcher captures vertical movement, and another Rotary Arm module on the same pitcher reflects pattern formation.

More broadly, this work suggests a larger research direction in passive non-contact detection, including passive distance detectors, airflow-responsive structures, and other physical systems that detect and communicate state change without conventional electronic sensing. At the same time, it raises a broader question for HCI's engagement with everyday life: if the value of many everyday objects lies partly in their passivity, how can we enrich them with more perceptual and interactive capacity without breaking that passive character? We see this as an important direction for future work, not only in developing new mechanisms, but also in understanding how passive objects can remain materially simple, unobtrusive, and compatible with everyday use while becoming more capable of sensing and expressing movement.


\section{Conclusion}

This paper introduced \name{}, a novel approach for designing passive mechanisms that respond to inertia-driven motion. We showed how passive objects can provide meaningful feedback about bodily movement while preserving the simplicity and low overhead that make them valuable in everyday life. We envision passive artifacts using their inherent physical properties to offer richer interaction and information, enabling people to build interactive objects while keeping with their passive nature.
\begin{acks}
Anonymous for review.
\end{acks}

\bibliographystyle{ACM-Reference-Format}
\bibliography{_references}

\appendix
\clearpage

\section{Experiment of Rotatory Arm}
\label{appendix:arm_experiment}

\begin{figure}[htbp]
    \centering
    \includegraphics[width=\linewidth]{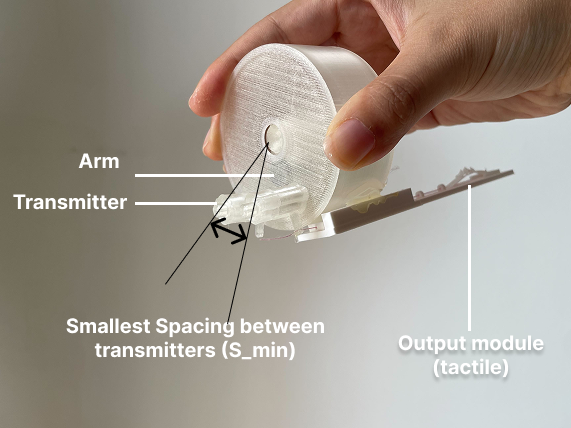}
    \caption{Experiment to find the minimum detectable angular step of the rotatory arm.}
    \label{fig:arm_experiment}
\end{figure}

The key dynamic variable of the rotatory arm is the minimum detectable angular step $\theta_{min}$, which represents the angular resolution of the structure. The rotatory arm detects rotational motion and triggers the transmitters connected to it when the rotatory arm sweeps across the circumference. When two transmitters are too close to each other, they would interact with each other and perturb the motion of the rotary arm, resulting in simultaneous activation even before the arm has reached them. The angle for two transmitters to start perturb each other is the $\theta_{min}$. To derive this $\theta_{min}$, we first measure the smallest spacing $S_{min}$ (which is the arc length) between the two transmitters. Then, using the relation between the arc length $S$, radius $R$ (which is the arm length of the rotatory arm), and angle $\theta$, we derive the dynamic threshold function to be
\begin{equation}
\begin{split}
    & S_{min} = R\theta_{min}\\
    & \theta_{min} = \frac{S_{min}}{R}
\label{eq: arc_radius}
\end{split}
\end{equation}

Figure \ref{fig:arm_experiment} illustrates our experiment setup to measure the smallest spacing $S_{min}$. The setup involves a rotatory arm structure with two transmitters attached, and an output module to detect the triggering of the two transmitters. The rotatory arm in our experiment can only hold a mass and a magnet, representing the minimum required length, as any shorter arm would be unable to support the mass and magnet necessary for the structure. The smallest spacing $S_{min}$ is constant. From Equation~\ref{eq: arc_radius}, for a fixed $S_{min}$, a smaller radius $R$ results in a larger $\theta_{min}$. Therefore, adopting the minimum arm length yields the largest achievable minimum detectable angular spacing of the structure. This minimum angular spacing can be reduced by increasing the arm length $R$. We then assemble the two transmitters onto this rotatory arm structure and gradually decrease the separation between the transmitters. Eventually, the separation between the two transmitters would start to perturb each other and the distance between the two transmitters at this point is $S_{min}$. We then use equation \ref{eq: arc_radius} to calculate $\theta_{min}$ from $S_{min}$.

\subsubsection{Minimum Detectable Angular Step Measured}
\label{appendix:arm_min_angle}
From the experiment, we measured the smallest spacing $S_{min}$ to be $15.95\ mm$, and the arm radius $R$ used in the experiment is $22.21\ mm$. Using equation \ref{eq: arc_radius}, the minimum detectable angular step $\theta_{min}$ is derived to be $0.718\ rad = 41.14^\circ$.

\section{Fabrication}
\label{appendix:fabrication}
\begin{figure}[!b]
    \centering
    \includegraphics[width=\linewidth]{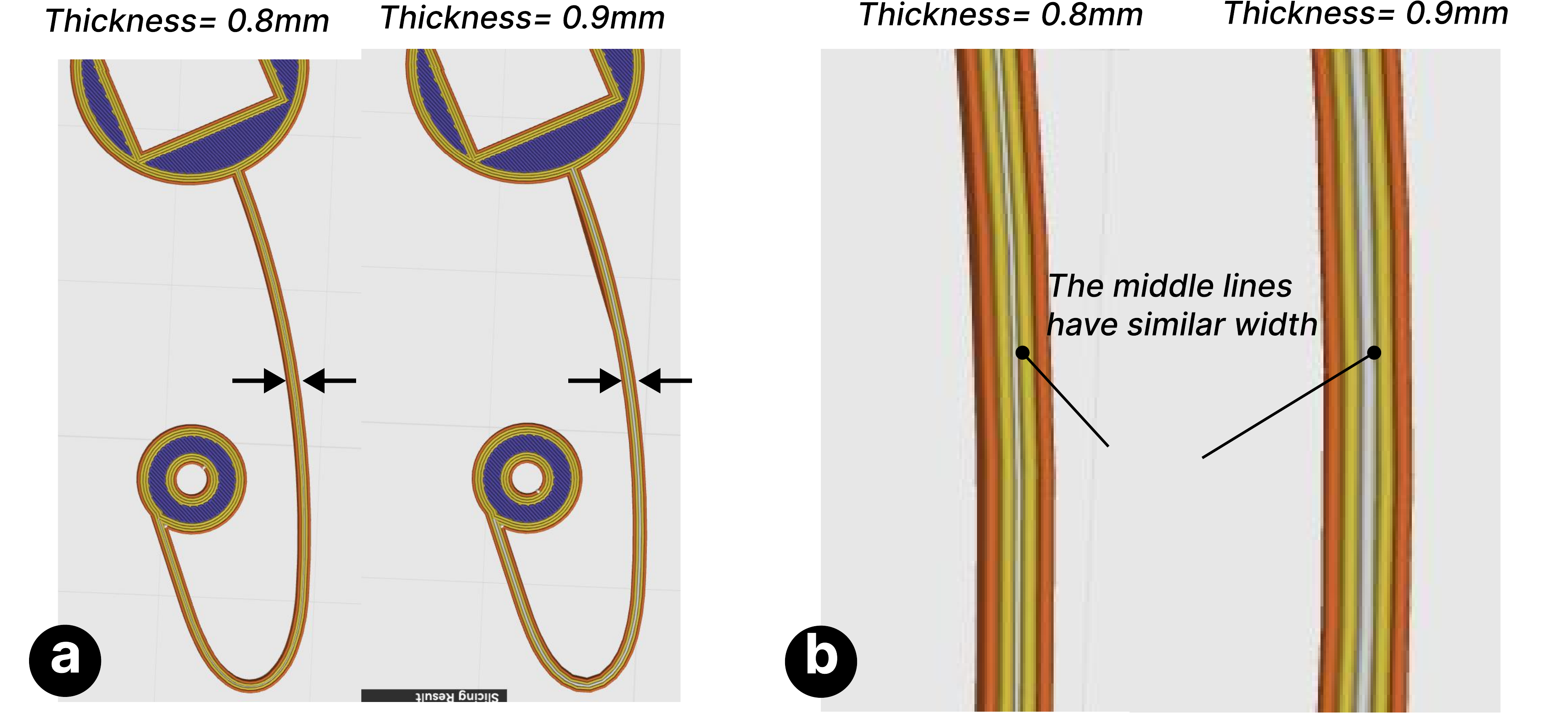}
    \caption{The sliced geometry in 3D-print slicer software.}
    \label{fig:fab}
\end{figure}
In this section, we describe how the primitives and modules were fabricated, along with practical issues encountered during testing and how we assemble them.
\subsection{FDM printing and Material}
Our prototypes were fabricated using FDM 3D printing with PETG and PLA. PETG was used for deformable parts, including the stopper pawl, cantilever beam, and tactile output, because it provides more flexibility while maintaining mechanical resistance. These parts were printed with a 0.2\,mm nozzle to better preserve small thickness differences in our parameterization.

However, small geometric differences did not always produce clearly distinguishable thresholding behavior. As shown in Figure \ref{fig:fab}, parts designed with only a 0.1\,mm difference sometimes performed similarly in threshold testing. This reflects a fabrication limitation: the printer could not reliably reproduce such small geometric differences. For this reason, we recommend prioritizing larger adjustments in length before relying on very small thickness changes to tune thresholds.

\begin{figure}[t]
    \centering
    \includegraphics[width=\linewidth]{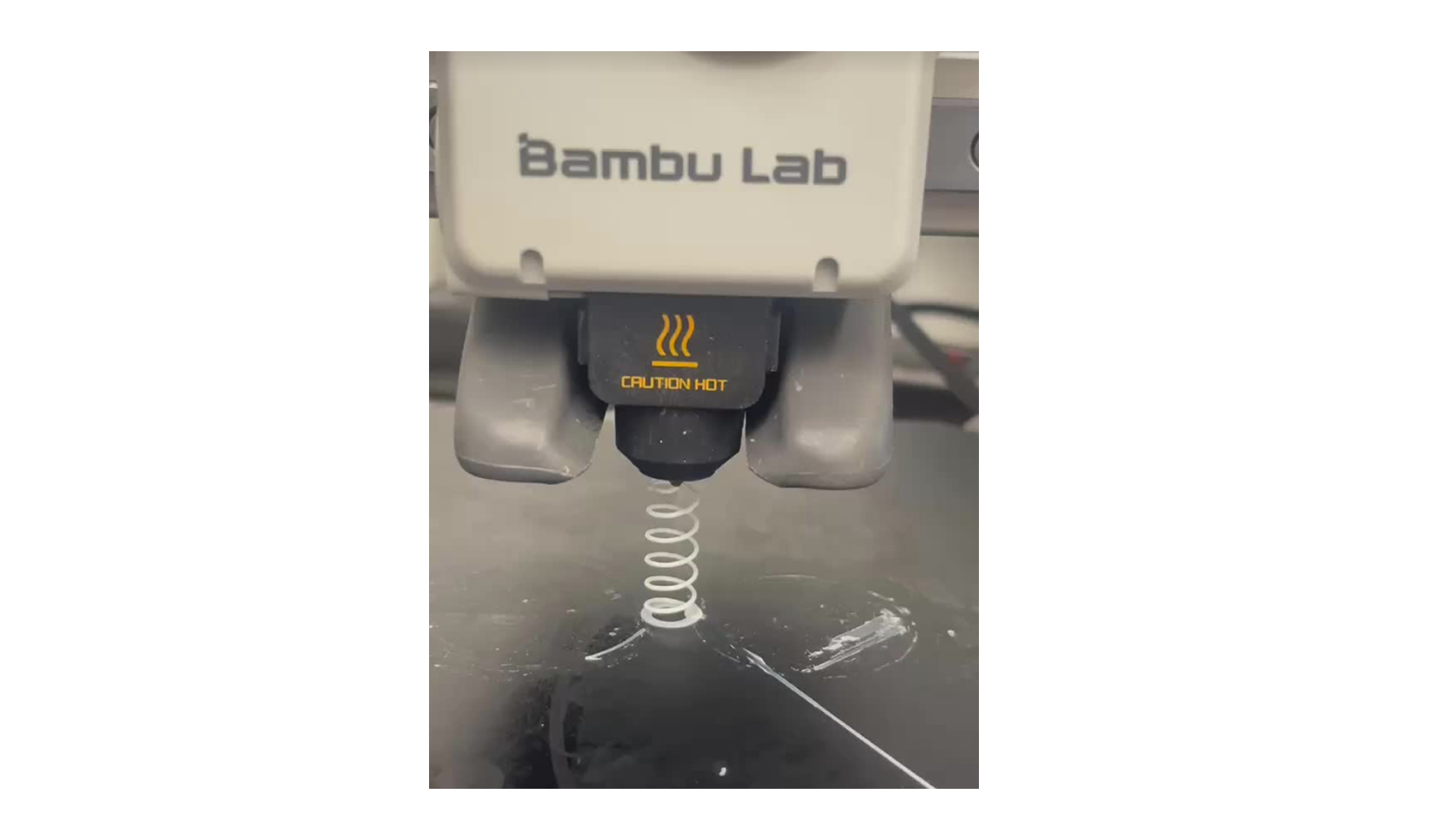}
    \caption{Print the spring through Bambu A1 mini Printer.}
    \label{fig:3D-print}
\end{figure}

The spring in the transmitter was also 3D printed in PLA (Figure \ref{fig:3D-print}), adapted from prior open-source work on printable springs \cite{mensley2018printsprings}.

\subsection{Assembling}
\begin{figure}[!b]
    \centering
    \includegraphics[width=\linewidth]{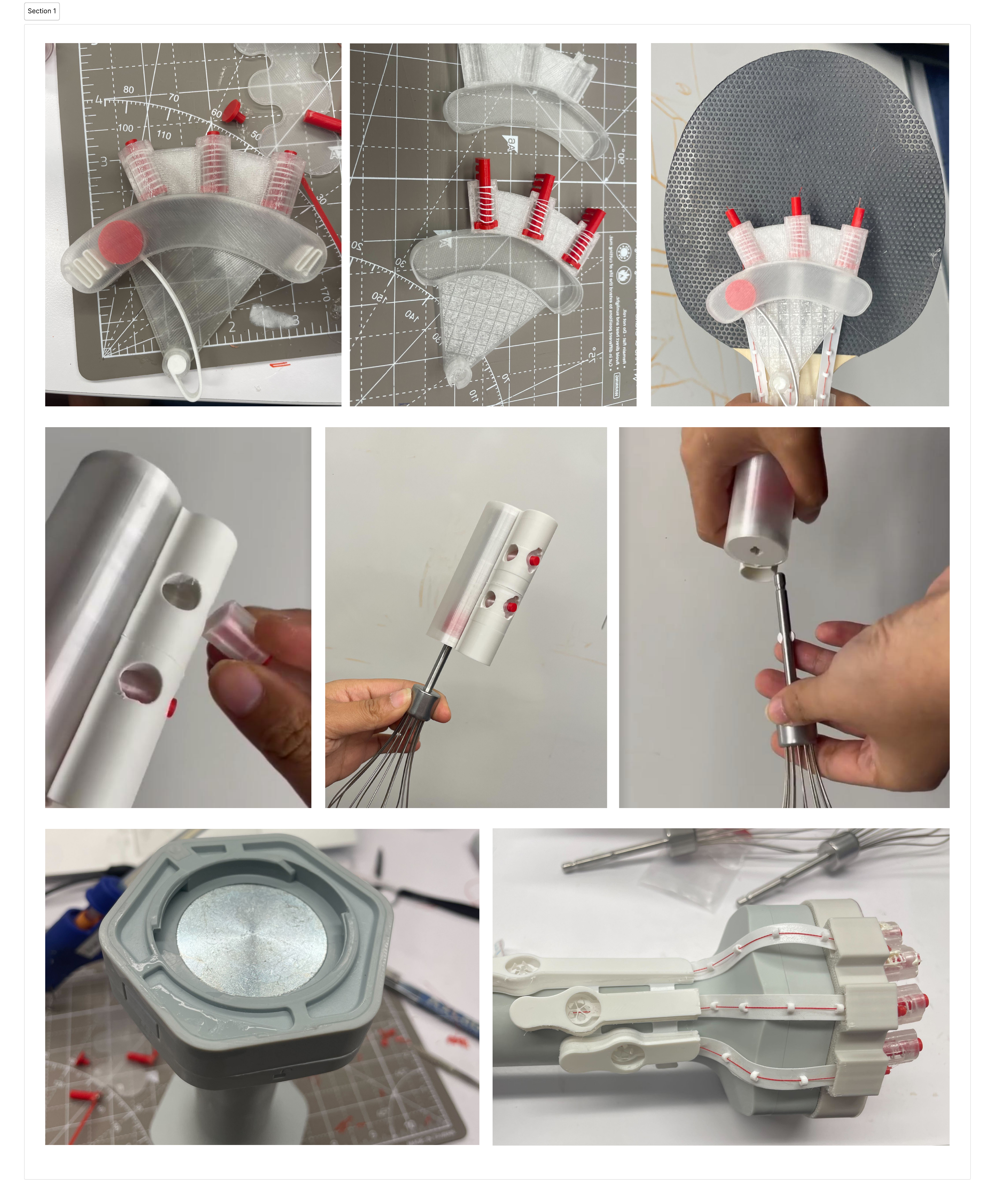}
    \caption{Assembling progress}
    \label{fig:assemble}
\end{figure}

We printed out all parts and assembled them together on the object. In this presented application case, we also modified the attached part where the our work connects to the object according to the object's form feature (Figure \ref{fig:assemble}). But most of the attachments are attached through gluing onto the object. 

\end{document}